\documentclass[journal]{IEEEtran}
\usepackage{amsmath,amsfonts,upgreek}
\usepackage{algorithmic}
\usepackage{algorithm}
\usepackage{array}
\usepackage[caption=false,font=normalsize,labelfont=sf,textfont=sf]{subfig}
\usepackage{textcomp}
\usepackage{stfloats}
\usepackage{url}
\usepackage{verbatim}
\usepackage{graphicx}
\usepackage{hyperref}
\usepackage{amssymb}
\usepackage{cancel}
\usepackage{booktabs}
\usepackage{colortbl}
\usepackage{balance}

\usepackage{multicol, multirow}
\usepackage{makecell}
\usepackage{xcolor}
\begin{document}

\title{A Variance-Preserving Interpolation Approach for Diffusion Models with Applications to Single Channel Speech Enhancement and Recognition}

\author{Zilu Guo, Qing Wang, Jun Du, Jia Pan, Qing-Feng Liu and Chin-Hui Lee,~\IEEEmembership{Life Fellow,~IEEE}
}

\markboth{IEEE/ACM TRANSACTIONS ON AUDIO, SPEECH, AND LANGUAGE PROCESSING}%
{Shell \MakeLowercase{\textit{et al.}}: A Sample Article Using IEEEtran.cls for IEEE Journals}


\maketitle

\begin{abstract}
In this paper, we propose a variance-preserving interpolation framework to improve diffusion models for single-channel speech enhancement (SE) and automatic speech recognition (ASR). This new \textcolor{black}{variance-preserving interpolation diffusion model (VPIDM)} approach requires only $25$ iterative steps and obviates the need for a corrector, an essential element in the existing variance-exploding interpolation diffusion model (VEIDM). Two notable distinctions between VPIDM and VEIDM are the scaling function of the mean of state variables and the constraint imposed on the variance relative to the mean's scale. We conduct a systematic exploration of the theoretical mechanism underlying VPIDM, and develop insights regarding VPIDM's applications in SE and ASR using VPIDM as a frontend. Our proposed approach, evaluated on two distinct data sets, demonstrates VPIDM's superior performances over conventional discriminative SE algorithms. Furthermore, we assess the performance of the proposed model under varying signal-to-noise ratio (SNR) levels. The investigation reveals VPIDM's improved robustness in target noise elimination when compared to VEIDM. Furthermore, utilizing the mid-outputs of both VPIDM and VEIDM results in enhanced ASR accuracies, thereby highlighting the practical efficacy of our proposed approach.
Code and audio examples are available online \footnote{\url{ https://github.com/zelokuo/VPIDM}}.

\end{abstract}


\section{Introduction}

\IEEEPARstart{A}{mbient} \textcolor{black}{noises, such as machine sounds, animal noises, and footsteps, are a common presence in our daily lives \cite{chen2006new}, and can impact the performance of automatic speech recognition (ASR) systems \cite{yu2016automatic,vincent_2016} and spoken question answering systems \cite{you2021knowledge, you2021self, you2022end, youCZ21, chenYZ21}. Speech enhancement (SE) technique \cite{boll1979suppression} aims to reduce noise while preserving the clarity of the speech signal, often approached as a supervised task \cite{wang2014training}, with deep learning (DL) methods being particularly successful, although unsupervised approaches are also being investigated \cite{li2021domain, lin2021unsupervised}.} SE has  Mask-based algorithms \cite{wang2014training, williamson_complex_2016}, similar to gain functions in traditional methods \cite{lim_oppen}, have been developed as learning targets. Another paradigm is the mapping function \cite{xu2014regression}, which transforms the noisy speech spectrum into a clean one. Beyond focusing on input and output targets, a wide variety of network architectures has been designed, including multi-layer perception (MLP) \cite{xu2014regression}, long short-term memory (LSTM) \cite{tu2020multi}, convolutional neural network (CNN) \cite{luo2019conv}, UNet \cite{tan_crn}, and Transformer \cite{cao22_interspeech}. Additionally, multi-target learning \cite{chen2015speech} and multi-stage models \cite{wu2006two} have been employed in SE tasks.
The concept of progressive learning (PL) has been proposed by researchers \cite{tu2020multi, gao2016snr}, involving a gradual noise removal process through the deepening layers of LSTM or MLP. Moreover,
regressive losses, such as minimize mean square error (MMSE) or minimum absolute error (MAE) \cite{wang2014training}, are crucial as cost functions for the DL-based SE task, leading to these methods being commonly referred to as regressive algorithms or discriminative algorithms.

Besides discriminative methods, generative models are also utilized for SE to estimate the distribution of clean speech. \textcolor{black}{Generative models are proposed based on the claim that the unconditional} distribution of clean \textcolor{black}{signals} is too complex to be directly represented by specific equations. However, this distribution can be implicitly modeled by artificial neural networks (ANNs). Variational autoencoders (VAE)  \cite{bie2022unsupervised} postulates that complex data distributions can be projected into a hidden state space via an encoder, where the hidden state variables conform to a multivariate Gaussian distribution. The clean data is then reconstructed by mapping this Gaussian state representation back to the real distribution using a decoder. Normalizing flows \cite{strauss2021flow} employs a sequence of invertible functions to transform a simple Gaussian distribution into the target distribution.
Generative adversarial networks (GAN) \cite{fu2019metricgan}, \cite{phan2020improving} use a discriminator to critique the generator, ensuring that the predicted clean speech closely resembles the real one. If the discriminator is omitted, the generator effectively becomes akin to a discriminative SE model. Therefore, following the study in \cite{richter_diff}, we still categorize GAN-based methods as discriminative approaches.

Diffusion models (DMs) have recently generated a large interest in SE \cite{richter_diff}, \cite{lu_apsipa, zhang21c_interspeech}, \cite{lu2022conditional, welker2022speech}, \cite{guo23_interspeech}, \cite{moliner_diff, srtnet}, \cite{cold_diff, storm_diff}, \cite{un_diff, rl_diff}, due to their success in various generative tasks, such as image generation \cite{song2019generative, ho2020denoising}, \cite{rombach2022high}, speech synthesis \cite{diff-wav}, and voice conversion \cite{popov2021diffusion}. Generative models seek a pathway from random noise to clean speech. Considering random noise and clean signal as starting and ending \textcolor{black}{states}, respectively, there exist numerous potential paths between them. Teaching an ANN to learn one of these paths can be challenging.
Intuitively, the diffusion process in  DMs can be seen as defining a series of \textcolor{black}{states}, guiding the reverse process in learning the path from random noise to clean speech. \textcolor{black}{In the context of discrete DMs, where the evolution between states is explicitly parameterized and the number of steps is finite, the process is governed by a parameter-free Markov chain. This ensures that each step is dependent on the immediate previous state, conforming to the definition of a Markov chain.} Specifically, in the diffusion process, each state variable in the chain is derived by incrementally adding Gaussian noise to the preceding state, starting from the clean data and gradually transitioning to a Gaussian distribution. Conversely, the reverse process involves progressively removing noise from each state variable, starting from Gaussian noise and converging to the clean signal.
The authors in \cite{song2020score} introduced the concept that state variables in discrete DMs are sampled from a continuous state space. They have formulated a stochastic ordinary differential equation (SODE) \cite{solin_2019} framework to model this state space, i.e., continuous DMs. Moreover, they pointed out that DMs described in \cite{ho2020denoising} and \cite{song2019generative} represent two specific cases of continuous DMs, termed variance-preserving diffusion model (VPDM) and variance-exploding diffusion model (VEDM), respectively. Furthermore, continuous DMs offer the advantage of smaller estimation errors, which is highly beneficial to DMs \cite{kingma2021variational}.

Adopting the conditional generation approach found in tasks such as text-to-image and text-to-speech, noisy speech can serve as a condition for generating clean speech in DMs \cite{lu_apsipa}, \textcolor{black}{but not achieving comparable results as discriminative models}. The condition extractor in \cite{lu_apsipa}  encodes noisy speech into a coarse-grained, high-level embedding, which inevitably discards significant fine-grained structures. However, this approach has not yet yielded top performances in SE, where fine-grained features play a crucial role in waveform reconstruction. \textcolor{black}{Another potential reason why directly applying DMs to SE tasks falls short is that original DMs are tailored for predicting distributions with greater flexibility than those required for SE. In contrast, each noisy speech clip matches only one clean clip, presenting a significant mismatch with the original design intent of DMs. Consequently, adapting DMs for SE tasks proves challenging and requires significant customization to align with the unique requirements of SE. }
The study in \cite{lu2022conditional} introduces the concept of incorporating noisy speech into the diffusion process to retain fine-grained information. Here, the mean of the state variable is a linear combination of clean and noisy speech, contrasting with the scaled clean speech in traditional DMs \cite{ho2020denoising, song2019generative}. Building on VEDM, authors in \cite{welker2022speech, richter_diff} extended the discrete state space in \cite{lu2022conditional} to a continuum and presented an SODE formula to model this space, leading to the development of a variance-exploding interpolation diffusion model (VEIDM). While VEIDM has achieved state-of-the-art (SOTA) performances, it may not efficiently enhance noisy speech in specific low Signal-to-Noise Ratio (SNR) conditions. Furthermore, VEIDM uses a corrector for improvement, it also doubles the number of the ANN's parameters, resulting in high computational costs.
Moreover, two-stage DMs have been proposed for SE \cite{srtnet, storm_diff}, wherein a discriminative model initially obtains a pre-enhanced signal, followed by a DM in the second stage to minimize the error between this pre-enhanced and clean speech, two steps that might lead to an increase in computational overheads.

These limitations of current DM-based SE algorithms have inspired us to develop a new interpolating scheme \cite{guo23_interspeech}.  Specifically, the mean of the state variable in our DM is essentially a linear interpolation of clean and noisy speech, scaled by a scheduling factor, where the mean constrains the variance. This proposed framework, termed variance-preserving interpolation diffusion model (VPIDM), has achieved SOTA performances among DM-based SE models. We believe we have two key contributions. In theory, we first establish a comprehensive principle, focusing on the methodology of the newly introduced VPIDM. Next, we perform an analysis of the impact of VPIDM on target noise reduction during the reverse DM process and propose an early-stopping rule to improve the robustness of ASR \cite{yu2016automatic,vincent_2016} of DM-enhanced speech using VPIDM as a frontend for ASR. In experiments, we evaluate the effectiveness of our proposed VPIDM on a large-scale data set, in contrast to previous DM-based SE studies, which predominantly focused on small-scale data sets \cite{lu2022conditional,richter_diff}. Moreover, we carry out extensive experiments to analyze the unique characteristics of VPIDM. For instance, we employ a critical discriminative baseline that shares the architectural backbone with both VPIDM and VEIDM, ensuring a fair comparative analysis.

\begin{table}[t]
    \color{black}
    \caption{List of Notations.}
    \label{tab_notions}
    \centering
    \resizebox{1.\columnwidth}{!}{%
    \begin{tabular}{ ll}
        \toprule
        {Symbol} &
        {Definition }
         \\
        \midrule
        $\mathbf{x},\mathbf{n}, \mathbf{y}$ & Clean speech, target noise, and noisy speech (time domain)   \\
        $ \mathbf{X}, \mathbf{N}, \mathbf{Y}$ & Clean speech, target noise, and noisy speech spectra    \\
        \midrule
        $\tau$ & State index, $0$ denotes the initial state, $T$ is the last state    \\
        $\mathbf{V}(\tau)$ & Linear interpolating process of  $\mathbf{X}$ and $\mathbf{N}$\\
    $ \lambda(\tau)$ & Interpolating coefficient, manipulating the ratio of  $ \mathbf{X}$ in $\mathbf{V}(\tau)$\\
    \midrule
$\mathbf{S}(\tau)$ & The state variable of forward or reverse process\\
$\mathbf{U}(\tau)$ & Mean vector of $\mathbf{S}(\tau)$, i.e., $\mathbf{S}(0)$ or $\mathbf{V}(\tau)$\\
$\mathbf{\Sigma}(\tau)$ &  Covariance matrix of $\mathbf{S}(\tau)$ \\
$\mathbf{Z}$ & Complex-valued circular symmetric Gaussian variable\\
$\alpha(\tau)$ & Scale coefficient, controlling the ratio of $\mathbf{U}(\tau)$ in $\mathbf{S}(\tau)$\\
$G(\tau)$ & SD coefficient, controlling the ratio of $\mathbf{Z}$ in $\mathbf{S}(\tau)$\\
$\mathcal{CN}$ & Complex standard norm distribution\\
\midrule
  $\mathbf{W}$ & Complex-valued Brownian motion in the forward process \\
   $\widetilde{\mathbf{W}}$ & Complex-valued Brownian motion in the reverse process \\
  $\text{d}(\cdot)$ & Differential operation \\
  $g(\tau)$ & Diffusion coefficients, representing the change rate of $\mathbf{\Sigma}(\tau)$ \\
   $\mathbf{f}(\mathbf{S}, \mathbf{Y}, \tau)$ & Drift coefficient, representing the change rate of $\mathbf{U}(\tau)$ \\
  $p(\mathbf{S}|\mathbf{X}, \mathbf{Y})$ & Conditional probability density of $\mathbf{S}(\tau)$ given  $ \mathbf{X}$ and $\mathbf{Y}$ pair \\
  $p_e(\mathbf{X}, \mathbf{Y})$ & Empirical joint probability of $ \mathbf{X}$ and $\mathbf{Y}$ pair in training set \\
  $\Psi_\theta(\mathbf{S}, \mathbf{Y}, \tau)$ & Output of the ANN \\
  $\frac{\text{d}(\cdot)}{\text{d}\mathbf{X}^*}$ & Complex gradient operation for a real function of $\mathbf{X}$ \cite{petersen2008matrix}\\
  $\mathbf{S}^{\textcolor{black}{q}}(\tau)$ & $q$-th $\mathbf{S}$ in a mini-batch, batch size is $Q$\\
  \midrule
   $\epsilon$ & Minimal state index (closest to the initial state in practice) \\
   $K$ & Number of states in a discrete state space \\
   $\Delta$ & $\frac{T - \epsilon}{K - 1}$ \\
   $\tau_k$ & $\tau_k = (k - 1)\Delta + \epsilon$ \\
  $\hat{\mathbf{S}}(\tau)$ & An estimation of $\mathbf{S}(\tau)$\\
  $\hat{\mathbf{V}}(\tau)$ & An estimation of $\mathbf{V}(\tau)$\\

  $\mathbf{S}_k$ & $k$-th state in a discrete state space sampled from $\mathbf{S}(\tau)$\\
  $\hat{\mathbf{S}}_k$ & An estimation of $\mathbf{S}_k$\\
  $\alpha_k$ & $\alpha(\tau_k)$ \\

        \bottomrule			
		\end{tabular} %
  }
	\end{table}

\section{Related work}
We first introduce the current SOTA DM-based SE algorithm, known as VEIDM. Moreover, we will also present VPDM, which primarily finds application in other fields. Nonetheless, it serves as the foundational concept and is essential for understanding our proposed VPIDM. We will use notations commonly used in \textcolor{black}{SE} literature to describe the signal models and related formulations which will be different from notations used in DM literature so far (e.g., \cite{lu_apsipa, richter_diff, welker2022speech}) and in our previous work \cite{guo23_interspeech}. \textcolor{black}{For convenience,
we summarize in Table \ref{tab_notions} the main notations used in this study.}

\subsection{Signal Model}
In this study, we consider the single-channel additive-noise signal model. Given a clip of clean speech $\mathbf{x}=[x_0, x_1, \cdots, x_{D-1}]^\intercal$ in the time domain and the additive noise $\mathbf{n}=[n_0, n_1, \cdots, n_{D-1}]^\intercal$, the resulting noisy speech $\mathbf{y}=[y_0, y_1, \cdots, y_{D-1}]^\intercal$ is given by
\begin{equation}
    \mathbf{y} = \mathbf{x} + \mathbf{n}.
    \label{time_mod}
\end{equation}
In this equation, the sets $\{\mathbf{x}, \mathbf{n}, \mathbf{y}\} \in \mathbb{R}^D$, and the signals are expressed in the time domain. Here, $D$ represents the total number of sample points, $x_d$, $n_d$, and $y_d$ represent the $d$-th element in $\mathbf{x}$, $\mathbf{n}$, and $\mathbf{y}$, respectively, $0 \le d\le D-1$, and $[\cdot]^\intercal$ signifies the  transpose of a vector or a matrix. To differentiate from the Gaussian noise in DMs, the $\mathbf{n}$ is termed as target noise in this article. Applying the short-time Fourier transform (STFT) to both sides of Eq. (\ref{time_mod}), we get:
\begin{equation}
    \mathbf{Y}^{tf} = \mathbf{X}^{tf} + \mathbf{N}^{tf}.
    \label{stft}
\end{equation}
Here, the superscript $tf$ denotes the signals are processed in the STFT domain, $\{\mathbf{X}^{tf}, \mathbf{N}^{tf}, \mathbf{Y}^{tf}\} \in \mathbb{C}^{L\times M}$,  $\mathbf{X}^{tf}=[X^{tf}_{l, m}]_{L\times M}$, $\mathbf{N}^{tf} = [N^{tf}_{l, m}]_{L\times M}$, and $\mathbf{Y}^{tf} = [Y^{tf}_{l, m}]_{L\times M}$ correspond to the STFT representations of the clean speech, target noise, and noisy speech, respectively. The $X^{tf}_{l, m}$, $N^{tf}_{l, m}$, and $Y^{tf}_{l, m}$, are the $l$-th row and $m$-th column elements in $\mathbf{X}^{tf}$, $\mathbf{N}^{tf}$, and $\mathbf{Y}^{tf}$, respectively. The indices $l$ and $m$ also denote the frame and frequency indices, with $l$ ranging from $0$ to $L-1$ and $m$ from $0$ to $M-1$.
To compensate for the typically heavy-tailed distribution of  \textcolor{black}{speech signal's STFT}, authors in \cite{welker2022speech, richter_diff} introduce a transformation that compresses the STFT spectrum into an amenable form for DM. It attempts to reduce the dynamic range of the \textcolor{black}{spectrum‘s complex value} without changing the phase spectrum of the original signal. The signal model is now represented by
\begin{align}
    \mathbf{Y} =  \mathcal{F}&(\mathbf{Y}^{tf}) = a|\mathbf{Y}^{tf}|^ce^{j\angle \mathbf{Y}^{tf}} ;
    \label{compress_function} \\
    \mathbf{Y} &\approx \mathbf{X} + \mathbf{N},
    \label{tf_model}
\end{align}
where $\mathbf{X} =  \mathcal{F}(\mathbf{X}^{tf})$, $\mathbf{N} =  \mathcal{F}(\mathbf{N}^{tf})$, $\mathcal{F}$ denotes the transformation function operating element-wise, $\mathbf{X}$, $\mathbf{N}$, and $\mathbf{Y}$ correspond to the transformed representations of the clean speech, target noise, and noisy speech, respectively. $j$ is the imaginary unit, $c \in (0, 1]$ is the transformed factor, $a\in (0, 1]$ manipulates the scale of the transformed signal,  and $|\cdot|$ represents the operation of computing the norm of a complex matrix element-wise, $[\cdot]^c$ represents all elements in the matrix to the $c$-th power, while $\angle\cdot$ signifies the phase of a complex matrix also computed element-wise.  The approximation in Eq. (\ref{tf_model}) is akin to the assumption made in \cite{boll1979suppression} that the magnitude of noisy speech is approximately equal to the sum of the magnitudes of the clean speech and the additive noise.
Moreover, the STFT operations described in the remainder of this paper will inherently include the transformation function as outlined in Eq.~(\ref{compress_function}).
Similarly, when conducting the inverse STFT (iSTFT), the procedure will commence with the application of the inverse function of $\mathcal{F}(\cdot)$.  Following the inverse function, the iSTFT is then performed to convert the frequency-domain signals back into their time-domain representations. Both VEIDM and our proposed VPIDM utilize the transformed representations as features.
\begin{figure}[tb]
  \centering
  \centering{\includegraphics[width=7.5cm]{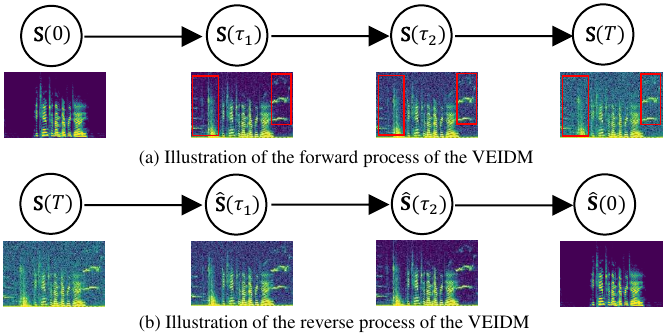}}
%
\caption{An illustration of the forward and reverse processes of VEIDM.
 }
\label{fig:veidm_foward}
\end{figure}

\subsection{VEIDM for Speech Enhancement}

VEIDM \cite{richter_diff} consists of two processes: the forward process and the reverse diffusion process, as illustrated in Fig. \ref{fig:veidm_foward}. For SE tasks, the two processes can be likened to analysis-synthesis methods \cite{analysis}.  The forward process resembles the analysis phase but lacks learnable parameters, \textcolor{black}{ which serve dual purposes: contributing to training the ANN and guiding the reverse process}. The reverse process parallels the synthesis stage, albeit in a recursive fashion.  \textcolor{black}{It is important to note that, unlike the forward process, the reverse process does not require clean speech to reconstruct the waveform.} VEIDM defines a bidirectional mapping from clean speech (starting state) to speech submerged in noises (ending state). During the forward, \textcolor{black}{ following the study \cite{song2020score}, VEIDM employs a continuous state space to present the process, which means there are countless middle states between the starting and ending states within the continum.} The states closer to the ending state are with more Gaussian noise and target noise. The reverse process learns to gradually remove a small portion of both noises in the guidance of the forward process \textcolor{black}{until obtaining clean speech estimation.} In addition, there are innumerable states between the starting and ending states, so we randomly select two states between the starting and ending to illustrate the forward process in Fig. \ref{fig:veidm_foward} (a).  We observe that the forward process is characterized by two key processes: the deterministic process and the stochastic process of gradually adding Gaussian noise.  The deterministic process is the linear interpolation of clean speech and noisy speech.
During the deterministic process which is indicated by the red rectangular in Fig. \ref{fig:veidm_foward} (a), noisy speech is incrementally merged with clean speech, leading to a gradual increase in the scale of the target noise. This process commences with clean speech and concludes when the noise scale closely matches that of the noisy speech.
Concurrently, in the stochastic process, Gaussian noise is incrementally introduced, leading to the deterministic part becoming submerged within the Gaussian noise.

\begin{figure}[t]
  \centering
  \centering{\includegraphics[width=7.5cm]{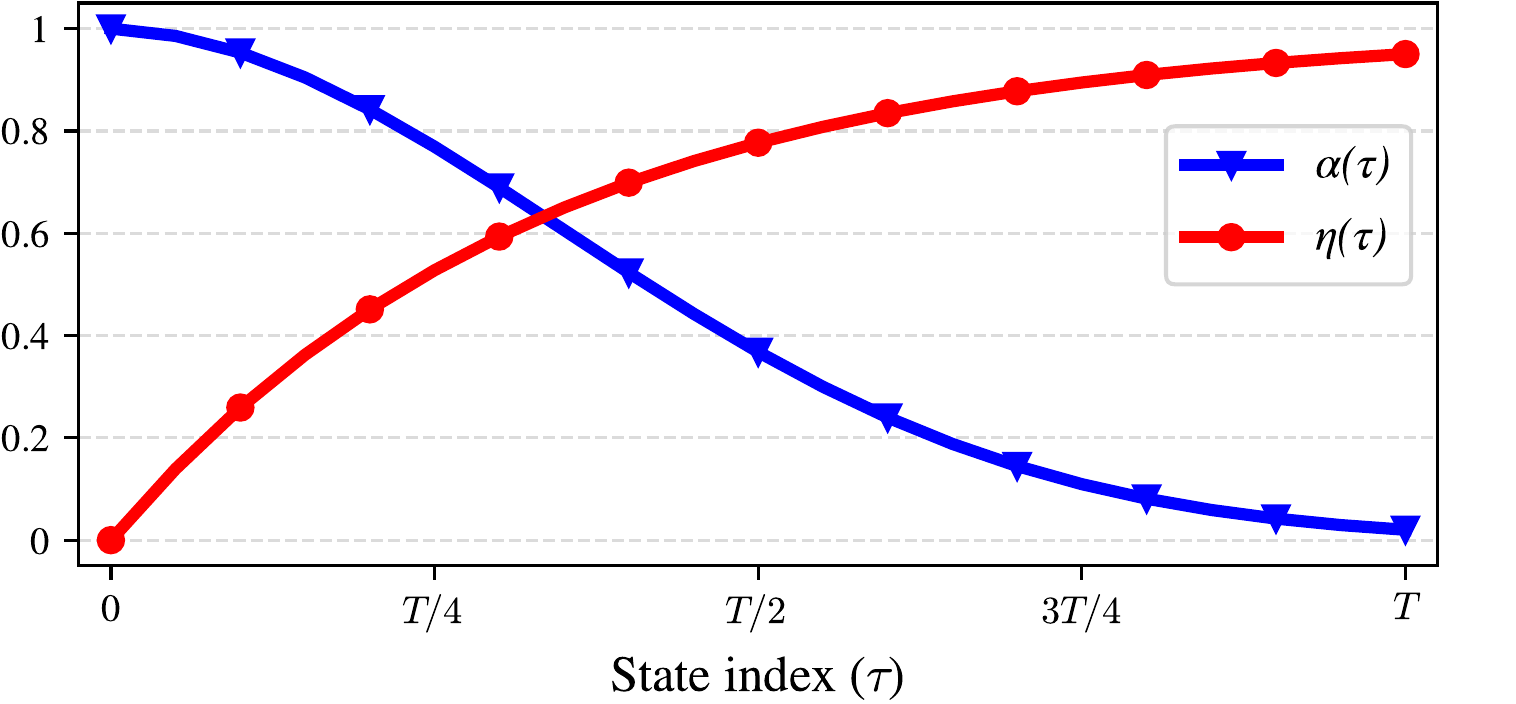}}
\caption{Two curves of $\eta(\tau)$ and $\alpha(\tau)$. $\eta(\tau)$ is the monotonically increasing function  w.r.t. $\tau$, and $\alpha(\tau)$ is the monotonically decreasing function. }
\label{fig:alpha_eta}
\end{figure}

In this study, we encapsulate the interpolation method \cite{richter_diff}, \cite{lu_apsipa,lu2022conditional}, \cite{welker2022speech} within a deterministic process represented by a state variable $\mathbf{V}(\tau)$ to provide a deeper understanding. To facilitate the definition of state variables in VEIDM, the three matrices  $\mathbf{X},  \mathbf{N}$ and $\mathbf{Y}$  are typically treated as three complex vectors, each belonging to the $LM$ dimension, i.e., $\{\mathbf{X}, \mathbf{N}, \mathbf{Y}\} \in \mathbb{C}^{LM}$ \cite{ho2020denoising, song2020score, richter_diff}.  This process assumes the target noise $\mathbf{N}$ is incrementally added to the clean speech $\mathbf{X}$ as the state index increases, resulting in the final state having a target-noise scale close to that of noisy speech.  The deterministic process is defined by
\begin{align}
     \mathbf{V}(\tau) &= \mathbf{X} + \eta(\tau)\mathbf{N}
     \label{add_noise_n} \\
        &= \lambda(\tau) \mathbf{X} + (1 - \lambda(\tau)) \mathbf{Y},
     \label{add_noise_y}
\end{align}
here $\mathbf{V}(0) = \mathbf{X}$ denotes the initial state (clean speech), $\eta(\cdot):\mathbb{R} \mapsto \mathbb{R}$ is a monotonically increasing function, termed the interpolating coefficient, $\eta(0) = 0$, and $\lambda(\tau) = 1 - \eta(\tau)$. An example curve of $\eta(\tau)$ is depicted in Fig. \ref{fig:alpha_eta}. When $\tau$ is from $0$ to $T$, the target-noise scale is gradually increased. Consequently, this process is termed the adding-target-noise process (ATNP).  Alternatively, $\mathbf{V}(\tau)$ can be the linear interpolation between clean and noisy speech shown in Eq. (\ref{add_noise_y}). During the reverse process, as $\tau$ decreases from $T$ to $0$, the intensity of the target noise diminishes, a phase we term the reducing-target-noise process (RTNP). During this process, the target noise in $\mathbf{V}(T)$ is gradually removed until $\mathbf{V}(\tau)$ degenerates to clean speech.
The stochastic process where the Gaussian noise is gradually added to the $\mathbf{V}(\tau)$  can be represented by
\begin{align}
    \mathbf{S}(\tau) &=\mathbf{V}(\tau) + G(\tau)\mathbf{Z},
    \label{veidm_state}
\end{align}
where $\mathbf{S}(\tau)$ is the state variable \textcolor{black}{in the continuous state space given state index $\tau$ (it is defined as a state variable in our study not solely based on its involvement in recursive updates, but also due to its fundamental role in representing the state space at any index $\tau$)}, $\mathbf{S}(0) = \mathbf{X}$ denotes the initial state (clean speech), $\tau$($0\le\tau\le T$) represents the state index, $T$ indicates the last state, $\mathbf{\Sigma}(\tau) = G^2(\tau)\mathbf{I}$ is the covariance matrix of $\mathbf{S}(\tau)$, $\mathbf{I}\in \mathbb{R}^{LM\times LM}$ is the unit diagonal matrix which means each element in $\mathbf{S}(\tau)$ is statistically independent, $G(\cdot):\mathbb{R} \mapsto \mathbb{R}$ is called the standard deviation (SD) coefficient, and $\mathbf{Z} \in \mathbb{C}^{LM}$ presents the complex-valued, circular symmetric Gaussian noise sampled from the complex standard norm distribution
\begin{equation}
    \mathbf{Z} \sim \mathcal{CN}(\mathbf{0}, \mathbf{I}),
    \label{gaussian_complex}
\end{equation}
here $\mathbf{0} \in \mathbb{R}^{LM}$ is a zero vector, and  $\mathcal{CN}$ denotes the complex standard norm distribution.
The reverse is illustrated in Fig. \ref{fig:veidm_foward} (b), which starts with the state in which both energies of Gaussian noise and target noise are high. Consequently, the Gaussian noise and target noise are gradually removed, until we get the estimation of clean speech. In addition,  the goal of the reverse is to estimate the clean speech, thus we can not resort to the state equation to predict the current state which is a function of the clean speech. In practice, the current state is recursively obtained from the previous state.

\subsection{VPDM for Generative Tasks}
VPDM was introduced in \cite{song2020score} for both unconditional and conditional image generation tasks. Here, to better understand our proposed VPIDM, we intend to modify VPDM \cite{song2020score} for SE by using noisy speech as a condition, aiming to estimate the distribution of clean speech. In VPDM, the state evolution equation is represented by
\begin{align}
			\mathbf{S}(\tau) = \alpha(\tau)\mathbf{S}(0) +\sqrt{1 - \alpha^2(\tau)} \mathbf{Z} \label{vp}.
			\end{align}
Here, the scale coefficient $\alpha(\cdot) :\mathbb{R} \mapsto \mathbb{R}$ is a monotonically decreasing function, with $\alpha(0)=1$, $0<\alpha(T)<1$. An illustrative curve of $\alpha(\tau)$ can be seen in Fig. \ref{fig:alpha_eta}. Moreover, the SD coefficient of Gaussian is constrained by the scale coefficient $\alpha(\tau)$. The ``variance-preserving" (VP) property in VPDM \cite{song2020score} states that the magnitude of the state variable is approximately unchanged during the whole forward process. Further detail of SODE for VPDM can be found in \cite{song2020score}.
While VPDM has demonstrated its superiority in other generative tasks, only few studies have achieved competitive performances in SE by directly using VPDM. Thus, we are encouraged to explore the possibility of enhancing VPDM.

\section{Proposed VPIDM for Speech Enhancement}
In this section, we first present our motivation. In addition, we will provide a detailed exposition of the state equation, drawing parallels to the formulations seen in Eqs. (\ref{veidm_state}) and (\ref{vp}) and also introduce SODE which is pivotal for understanding the reverse process.
Accordingly, we will articulate the training process, and define the training target. We next delve into the reverse process to showcase the procedure of enhancing a clip of noisy speech. We finally present an interpretation of our proposed VPIDM, offering some insights and analytical perspectives into comprehending the implications of the proposed approach in sub-sections (\ref{sub_veidm_vpidm_vpdm}), (\ref{reducing_noise_reason}), and (\ref{subsec_asr}).

\subsection{Motivation}
The current VEIDM \cite{richter_diff} has achieved SOTA performance for SE by utilizing a powerful ANN model. However, during the reverse process, the ANN's estimation at every step is not so accurate and needs to be enhanced by a corrector \cite{song2020score, richter_diff}. The corrector also re-implements the ANN as the backbone. Consequently, it leads to the total inferring time doubling. Moreover, we find that the performance of VEIDM can not transcend the discriminative model which uses the same ANN model as the backbone when the corrector is muted.  Furthermore, for the initial state $\mathbf{S}(T)$ (in Eq. (\ref{veidm_state})) of the reverse process, the clean speech is \textcolor{black}{unavailable}, thus an approximate $\mathbf{S}(T)$ is obtained by replacing the clean speech with the noisy speech in practice. Therefore, it is inevitable to cause the error, termed the initial error.  Notably, we can add more Gaussian noise to each state for VEIDM to obtain a relatively small initial error. However, this strategy will cause the scale range of $\mathbf{S}(\tau)$ to increase which is detrimental for the ANN to learn the training target. It is reported that the corrector is not necessary for the VPDM \cite{song2020score}. Besides, the VP strategy could reduce the initial error as we point out in our previous paper \cite{guo23_interspeech}.
\begin{figure}[t]
  \centering
  \centering{\includegraphics[width=8.5cm]{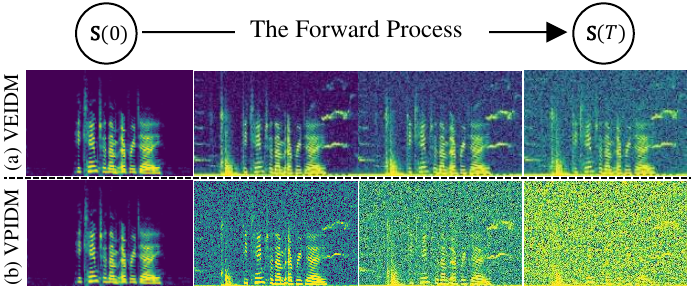}}
%
\caption{A comparison of forward processes of VEIDM and VPIDM, with four logarithmic spectra sampled from the respective processes.}
\label{fig:vpidm_veidm_illustration}
\end{figure}
Therefore, it is encouraged for us to apply the VP strategy to VEIDM or adopt the interpolation method for VPDM in the context of SE.  Consequently, we tailor the VEIDM to the proposed VPIDM. As depicted in Fig. \ref{fig:vpidm_veidm_illustration} (a), the deterministic components account for a considerable portion compared to the stochastic Gaussian noise in $\mathbf{S}(T)$ for VEIDM, thus the initial error could have detrimental impacts on the reverse process.  In Fig. \ref{fig:vpidm_veidm_illustration} (b), we illustrate the diffusion process of the proposed VPIDM. We observe that the deterministic components are submerged into the Gaussian noise which thereby could cause less initial error.

\subsection{Extending VPDM to VPIDM for Speech Enhancement}
\label{sec_VPIDM}
\label{signal_model}
In line with the methodologies discussed in \cite{ho2020denoising, song2020score}, we assume that the state equation in the forward diffusion process is an affine function, composed of a deterministic and a stochastic Gaussian component. Furthermore, drawing inspiration from \cite{welker2022speech, richter_diff, song2020score}, we propose that the mean itself is an affine function of both clean and noisy speech, as defined in Eq. (\ref{add_noise_y}). On top of the new state equation, we also express SODE in a closed form for VPIDM in this sub-section.
\subsubsection{The Forward Process}
The forward process comprises two crucial equations: the state equation, which establishes the connection with the initial state (clean speech), and the SODE, which serves as the crucial foundation for the reverse process.
The state equation is crucial for sampling at any given state index $\tau \in [0, T]$, relative to the clean-noisy speech pair.
The state equation of VPIDM is represented by
\begin{align}
    \mathbf{S}(\tau) &\triangleq \alpha(\tau)\left[\lambda(\tau)\mathbf{X} + (1 -\lambda(\tau))\mathbf{Y}\right]+ \sqrt{1-\alpha^2(\tau)}\mathbf{Z}  \nonumber\\
    &= \alpha(\tau)\mathbf{V}(\tau) + \sqrt{1-\alpha^2(\tau)}\mathbf{Z},
    \label{IDM_se}
\end{align}
here the scale coefficient $\alpha(\tau)$ has a similar form to that in Eq. (\ref{vp}), and $\lambda(\tau) = 1 - \eta(\tau)$ is the same as that in Eq. (\ref{add_noise_y}).
The state variable $\mathbf{S}(\tau)$ and $\mathbf{V}(\tau)$ as defined in Eq. (\ref{IDM_se}),
are central to this model. The initial state of the model is set as $\mathbf{S}(0) = \mathbf{V}(0) = \mathbf{X}$, the clean signal.
The forward SODE of VPIDM (and also VEIDM detailed in \cite{richter_diff}) is given by the unified framework proposed in \cite{song2020score}:
\begin{equation}
	\text{d}\mathbf{S}(\tau) = \mathbf{f}(\mathbf{S}, \mathbf{Y}, \tau)\text{d}\tau + g(\tau)\text{d}\mathbf{W},
                \label{idmd}
\end{equation}
where $\mathbf{f}(\cdot, \mathbf{Y}, \tau): \mathbb{C}^{LM} \mapsto \mathbb{C}^{LM}$, and $g(\cdot): \mathbb{R} \mapsto \mathbb{R}$ are the drift and diffusion coefficients, respectively,
and $\mathbf{W}$ is the complex-valued Brownian motion. Suppose the increment of $\tau$ is $\Delta (\rightarrow 0)$, then $\Delta\mathbf{W} \sim \mathcal{CN}(\mathbf{0}, \Delta \mathbf{I})$. Notably, VEIDM and VPIDM follow the same unified SODE as presented in Eq. (\ref{idmd}), but with discrepant coefficients. According to Eqs. (5-50) and (5-51) in \cite{solin_2019}, the two coefficients in VPIDM are:
\begin{align}
    \mathbf{f}(\mathbf{S}, \mathbf{Y}, \tau) &= \frac{\text{d}\ln\left[{\alpha(\tau) \lambda(\tau)}\right]}{\text{d}\tau}\mathbf{S}(\tau) - \alpha(\tau)\frac{\text{d}\ln\lambda(\tau)}{\text{d}\tau}\mathbf{Y}
    \label{drift_co_idm};  \\
 g(\tau) &= \sqrt{\frac{\text{d}G^2(\tau)}{\text{d}\tau} - 2G^2(\tau) \frac{\text{d}\ln\left[{\alpha(\tau) \lambda(\tau)}\right]}{\text{d}\tau}}.
    \label{diff_co_idm}
\end{align}
A detailed derivation of \textcolor{black}{the} two above coefficients is formulated in Eqs. (\ref{derivation_of_drift}) and (\ref{derivation_diffusion}) shown in Appendix. We follow \cite{welker2022speech, richter_diff}, and express $\lambda(\tau)$ in an exponential form and set the scale coefficient $\alpha(\tau)$ in an identical form to that in \cite{song2020score}, but with customized hyper-parameters for SE:
\begin{align}
    \lambda(\tau) &=  e^{-\gamma\tau};
    \label{vpidm_lamda} \\
    \alpha(\tau) &= e^{-0.5\int_0^\tau \beta(s) \text{d}s};
    \label{vpidm_alpha}\\
    G(\tau)  &= \sqrt{1 - \alpha^2(\tau)},
    \label{vpidm_G}
\end{align}
where the non-negative hyper-parameter $\gamma$ manipulates the speed of infusing noisy speech, and $\beta(\tau) = (\beta_\text{max} - \beta_\text{min})\tau + \beta_\text{min}$, with two non-negative constant hyper-parameters, $\beta_{\text{max}}$ and
 $\beta_{\text{min}}$, controls the rate of change from $\mathbf{S}(0)$ to $\mathbf{S}(T)$.
 \begin{figure}[b]
  \centering
  \centering{\includegraphics[width=8.5cm]{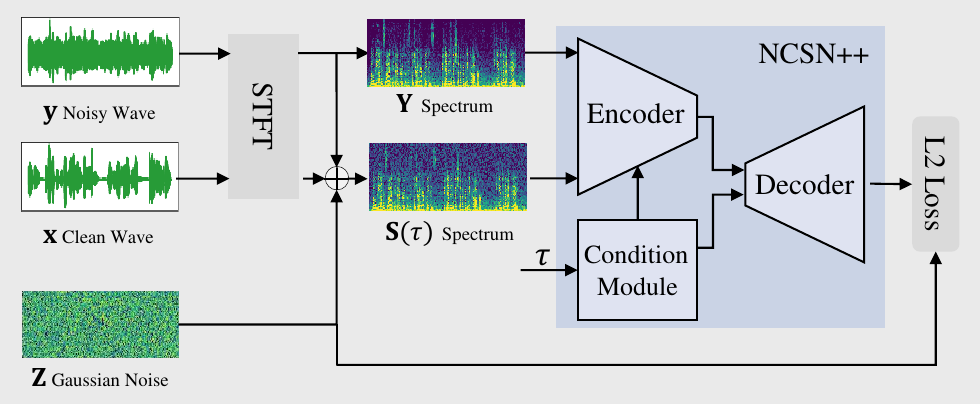}}
\caption{An illustration of the training stage. The spectrum of $\mathbf{S}(\tau)$, the noisy spectrum, and the state index are injected into the ANN to predict the weighted Gaussian noise in $\mathbf{S}(\tau)$. The L2 loss is utilized as the cost function.}
\label{fig:tranining_process}
\end{figure}
\begin{figure*}[htb]
  \centering
  \centering{\includegraphics[width=18cm]{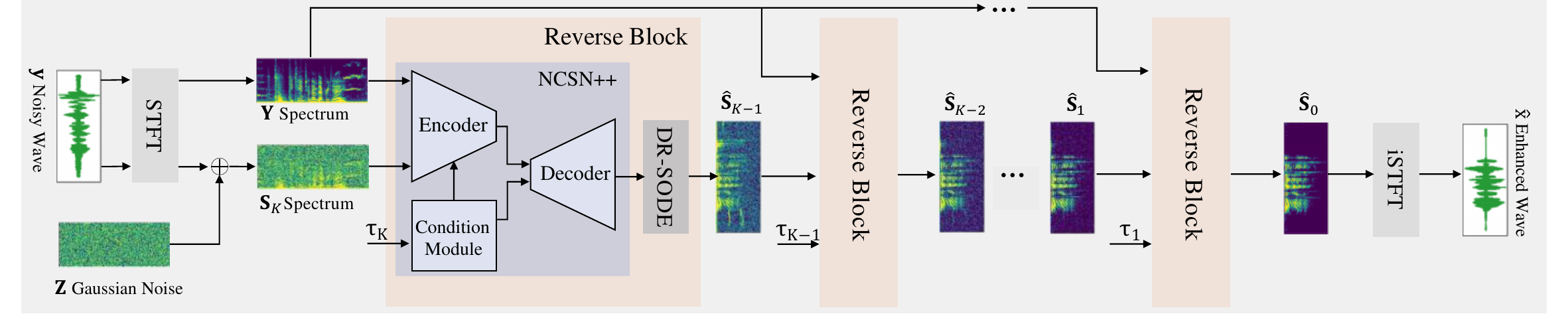}}
%
\caption{A diagram of the reverse process. The initial state variable $\mathbf{S}_K$ sampled from Eq. (\ref{initial_reverse}) with the noisy spectrum and the state index is fed into the Reverse Block to get an estimation of the next state $\mathbf{S}_{K-1}$.
Re-utilize the Reverse Block $K$ times, we finally get an estimation of the clean speech spectrum, then transform it to the clean waveform via iSTFT. The Reverse Block consists of two modules: ANN and DR-SODE denoted in Eq. (\ref{reverse_recursive_equation_discrete}). }
\label{fig:diffusion_process_nn}
\end{figure*}
\subsubsection{Training Target and Loss}
Training is illustrated in Fig. \ref{fig:tranining_process}. By inputting clean and noisy speech to \textcolor{black}{evaluate} STFT, we get clean and noisy speech spectra. In addition, we randomly select $\tau$ and sample $\mathbf{Z}$ from Eq. (\ref{gaussian_complex}). We now obtain the state variable $\mathbf{S}(\tau)$ from Eq. (\ref{IDM_se}) and then input the state variable, noisy spectrum, and the state index to NCSN++ (introduced in Section \ref{neural_network}). The loss is computed by Eq. (\ref{loss_prctice}) to be presented later.
It is worth noting that $\mathbf{S}(\tau)$ requires a pair of clean and noisy speech for supervision. The ANN output, $\Psi_\theta(\mathbf{S}, \mathbf{Y}, \tau) \in \mathbb{C}^{LM}$ where the subscript $\theta$ denotes the parameters of the ANN, is to predict the \textit{score} \cite{vincent2011connection, song2019generative, song2020score} which is represented in the complex domain in this paper. It is denoted as $\frac{\text{d}\ln p(\mathbf{X}|\mathbf{Y})}{\text{d}\mathbf{X}^*}$, where superscript $*$ signifies the conjugate operation, $p(\mathbf{X}|\mathbf{Y})$ represents the conditional probability density function of clean speech given noisy speech. Unlike VAEs which seek to predict the conditional probability $p(\mathbf{X}|\mathbf{Y})$, DMs attempt to estimate the score, i.e., gradient of the conditional probability, to avoid the problem of intractable normalization term \cite{song2019generative}. The training objective is to minimize MMSE of $\Psi_\theta$ and the score $\frac{\text{d}\ln p(\mathbf{X}|\mathbf{Y})}{\text{d}\mathbf{X}^*}$ as follows,
\begin{equation}
\label{orignal_loss}
    \underset{\theta}{\text{arg min}}\mathbb{E}_{p(\mathbf{X}|\mathbf{Y})}\left|\left|\Psi_\theta(\mathbf{S}, \mathbf{Y}, \tau) -\frac{\text{d}\ln p(\mathbf{X}|\mathbf{Y})}{\text{d}\mathbf{X}^*} \right|\right|^2_2,
\end{equation}
where $||\cdot||_2^2$ denotes the square of the L$2$ norm, and  $\mathbb{E}$ signifies the mathematical expectation.
However, the \textcolor{black}{unavailable} conditional probability makes the score in Eq.~(\ref{orignal_loss}) \textcolor{black}{inaccessible}, \textcolor{black}{causing} the expectation in Eq.~(\ref{orignal_loss}) impracticable. According to denoising score-matching in \cite{vincent_2016, song2020score}, the optimization problem presented in Eq. (\ref{orignal_loss}) becomes:
\begin{align}
  &\underset{\theta}{\text{arg min}}\mathbb{E}_{p(\mathbf{S}|\mathbf{X}, \mathbf{Y})p_e(\mathbf{X}, \mathbf{Y})}\left|\left|\Psi_\theta(\mathbf{S}, \mathbf{Y}, \tau) - \frac{\text{d}\ln p(\mathbf{S}|\mathbf{X}, \mathbf{Y})}{\text{d}\mathbf{S}^*}\right|\right|^2_2 \nonumber \\
      &= \underset{\theta}{\text{arg min}}\mathbb{E}_{p(\mathbf{S}|\mathbf{X}, \mathbf{Y})p_e(\mathbf{X}, \mathbf{Y})}\left|\left|\Psi_\theta(\mathbf{S}, \mathbf{Y}, \tau) +\frac{\mathbf{Z}}{G(\tau)} \right|\right|_2^2
    \label{score_match}
\end{align}
where $p(\mathbf{S}|\mathbf{X}, \mathbf{Y})$ represents the conditional density of $\mathbf{S}(\tau)$ given clean and noisy speech, $p_e(\mathbf{X}, \mathbf{Y})$ denotes the empirical joint density of clean and noisy speech given the training set. The conditional density $p(\mathbf{S}|\mathbf{X}, \mathbf{Y})$ and its gradient are:
\begin{align}
 \mathbf{U}(\tau) = \mathbb{E}_{\mathbf{S}}\left[\mathbf{S}\right] 
    &=\alpha(\tau)\left[\lambda(\tau)\mathbf{X} + (1- \lambda(\tau))\mathbf{Y}\right];
    \label{mean} \\
     p(\mathbf{S}|\mathbf{X}, \mathbf{Y}) ~
    &= ~\frac{1}{(\pi)^{LM}\det(\mathbf{\Sigma})}e^{-\left(\mathbf{S} - \mathbf{U}\right)^\text{H}\mathbf{\Sigma}^{-1}\left(\mathbf{S} - \mathbf{U}\right)}; \label{conditional_prob}\\
      \frac{\text{d}\ln p(\mathbf{S}|\mathbf{X}, \mathbf{Y})}{\text{d}\mathbf{S}^*}
    &= -\frac{\left(\mathbf{S} - \mathbf{U}\right)}{G^2(\tau)} = - \frac{\mathbf{Z}}{G(\tau)}.
    \label{score_condition}
\end{align}
Here $\det(\mathbf{\Sigma})$ is the determinant of  the covariance matrix $\mathbf{\Sigma}(\tau)$, the superscript $\text{H}$ denotes conjugate (or Hermitian) transpose. In practice,  using a Monte Carlo method \cite{kingma2021variational} to  approximate the expectation in Eq. (\ref{score_match}) in a mini-batch, and weighting it with the SD coefficient $G(\tau)$ to keep training stable \cite{song2019generative, song2020score},  the cost function is now represented as:
\begin{equation}
    \mathcal{L} = \sum^{\textcolor{black}{Q}}_{q=1}\frac{\left|\left|G\left(\tau^{\textcolor{black}{q}}\right)\Psi_\theta\left[\mathbf{S}^{\textcolor{black}{q}}, \mathbf{Y}^{\textcolor{black}{q}}, \tau^{\textcolor{black}{q}}\right] +\mathbf{Z}^{\textcolor{black}{q}}\right|\right|_2^2}{QLM},
    \label{loss_prctice}
\end{equation}
where $Q$ is the batch size, the superscript $q$ denotes the $q$-th signal in the batch, $\tau$ is uniformly sampled from $(\epsilon, T]$. $\epsilon$ presents the minimal state index indicating the first state after the clean signal during the diffusion process, an important hyper-parameter impacting the training stability \cite{song2020score, guo23_interspeech}.

\subsubsection{The Reverse Process} Fig. \ref{fig:diffusion_process_nn} shows an illustration of the reverse process where we reuse the Reverse Block $K$ times but input different state variables and indices into the block to obtain the estimated clean spectrum recursively. We first sample an initial state $\mathbf{S}_K$ and input it with the noisy spectrum and the state index to get an estimate of the next state. By repeating this step $K$ times, we finally get the enhanced speech spectrum.  Then, we apply iSTFT to get enhanced speech. Next, we will provide a detailed explanation of the meanings of $\mathbf{S}_K$, $K$, and the reverse process.
The reverse SODE of our proposed VPIDM akin to those \cite{vincent2011connection, song2020score} is defined as:
\begin{equation}
    \text{d}\mathbf{S}(\tau)\!=\!\!\left[ - \mathbf{f}(\mathbf{S},\!\mathbf{Y},\!\tau) \!+ \!g^2(\tau) \frac{\text{d}\ln p(\mathbf{X}|\mathbf{Y})}{\text{d}\mathbf{X}^*} \right]\!\text{d}\tau\! +\! g(\tau)\text{d}\widetilde{\mathbf{W}}
    \label{idm_re}
\end{equation}
where $\widetilde{\mathbf{W}}$ is another complex-valued Brownian motion independent from $\mathbf{W}$, $\mathbf{f}(\mathbf{S}, \mathbf{Y}, \tau)$ and $g(\tau)$  are the drift and diffusion coefficients defined in Eqs. (\ref{drift_co_idm}) and (\ref{diff_co_idm}), respectively. Replacing $\frac{\text{d}\ln p(\mathbf{X}|\mathbf{Y})}{\text{d}\mathbf{X}^*}$ with the ANN output, we get:
\begin{equation}
    \text{d}\mathbf{S}(\tau) \!= \! \left[-\mathbf{f}(\mathbf{S},\!\mathbf{Y},\!\tau) \!+ \!g^2(\tau) \Psi_\theta(\mathbf{S},\!\mathbf{Y},\!\tau) \right] \text{d}\tau \!+\! g(\tau)\text{d}\widetilde{\mathbf{W}}.
    \label{idm_re_nn}
\end{equation}
In utilizing R-SODE in Eq. (\ref{idm_re_nn}) to obtain an estimate of the clean speech spectrum, we have to calculate the entire reverse process which is computationally demanding. Therefore, a discrete R-SODE (DR-SODE) is adopted in practice as shown in Eq. (\ref{reverse_recursive_equation_discrete}).
Dividing $[\epsilon, T]$ evenly into $K - 1$ equal parts, $K$ is the total sampling steps, $\Delta=\frac{T - \epsilon}{K - 1}$.  R-SODE and discrete functions within R-SODE are defined by
\begin{align}
\mathbf{S}_{k-1}  = \mathbf{S}_{k}\!-\![\mathbf{f}_k(\mathbf{S}_k, \mathbf{Y}) - & g^2_k \Psi_{\theta,k}(\mathbf{S}_k,\!\mathbf{Y})] \Delta + g_k\sqrt{\Delta}\mathbf{Z} \label{reverse_recursive_equation_discrete} \\
    \mathbf{f}_{k }(\mathbf{S}_k, \mathbf{Y}) &= \mathbf{f}(\mathbf{S}(\tau_k), \mathbf{Y},\tau_k);  \\
      \Psi_{\theta, k}(\mathbf{S}_k, \mathbf{Y}) &= \Psi_\theta\left(\mathbf{S}(\tau_k), \mathbf{Y},\tau_k \right);  \\
       \mathbf{\psi}_k =  \mathbf{\psi}(\tau_k), ~~ &\text{for } \mathbf{\psi} \in \{\mathbf{S},\mathbf{V}, \mathbf{U}\};  \\
        \rho_k =  \rho(\tau_k), ~~ &\text{for } \rho \in \{\alpha, \eta, \lambda, G, g\}; \\
        \tau_k = (k - 1)\Delta + \epsilon, &~\text{for } k \in \{ 1, 2, 3, \cdots, K\}.
\end{align}
During the reverse diffusion process,  starting with $\mathbf{S}_K$ sampled from Eq. (\ref{IDM_se}) is impracticable, because the clean signal is also required. In practice, we replace the clean signal with the noisy spectrum \cite{lu2022conditional, richter_diff} when $k = K$ and hence sample $\mathbf{S}_K$
\begin{equation}
    \mathbf{S}_K \sim \mathcal{CN}\left(\alpha_K\mathbf{Y}, G^2_K\mathbf{I}\right).
    \label{initial_reverse}
\end{equation}

\subsection{VPIDM in Contrast to VEIDM and VPDM}
\label{sub_veidm_vpidm_vpdm}

The general  state equation  related to VEIDM and VPIDM can be expressed as:
\begin{equation}
			\mathbf{S}(\tau) = \alpha(\tau)\mathbf{V}(\tau)  +G(\tau) \mathbf{Z} \label{idm}.
\end{equation}
This framework states that the deterministic process is constrained by the scale coefficient $\alpha(\tau)$ and the interpolating factor $1 - \lambda(\tau)$ in $\mathbf{V}(\tau)$. They might operate independently of the stochastic Gaussian process which is controlled by the SD coefficient $G(\tau)$. We list the similarities and differences between VEIDM and VPIDM in Table \ref{table_idm}.
It's clear that VEIDM  and VPIDM  are two distinct variants of Eq. (\ref{idm}).
In VEIDM, the value of $\alpha(\tau)$ is consistently set to $1$, whereas VPIDM constrains the variance of the Gaussian process to be related to the scale coefficient $\alpha(\tau)$.

According to Section \ref{sec_VPIDM}, different state equations induce discrepant reverse processes. From state equations in Eqs. (\ref{IDM_se}) and (\ref{veidm_state}), we observe that speech components gradually diminish in VPIDM, and keep unchanged in VEIDM during the forward process. As a result, during the reverse process, VPIDM reconstructs clean speech's amplitude progressively, whereas VEIDM argues that the amplitude is unchanged during the whole reverse process because $\alpha(\tau)$ constantly equals $1$ (we will also present the analysis in Section \ref{reducing_noise_reason}). Reconstructing clean speech's amplitude progressively renders the ANN to learn the small changes between two states, providing more rich information than VEIDM.
\begin{table}[t]
		\caption{A Comparison of State Equations for VEIDM/VPDM/VPIDM.}

		\label{table_idm}
		\centering
		\begin{tabular}{ cccc}
			\toprule
			   {\makecell[c]{Diffusion \\ Models}}  &
			  {\makecell[c]{Scale\\ Coefficients}} &
			  {\makecell[c]{SD\\ Coefficients}}  &
             {\makecell[c]{Interpolating \\Coefficients}}\\
			\midrule    	
      VEIDM \cite{richter_diff}  &   1 &   $G(\tau)$ & $1 - \lambda(\tau)$ \\
       VPDM  &    $\alpha(\tau)$ &   $\sqrt{1 - \alpha^2(\tau)}$ & $0$  		\\
       VPIDM  &  $\alpha(\tau)$ &   $\sqrt{1 - \alpha^2(\tau)}$ & $1 - \lambda(\tau)$ \\
			
			\bottomrule
		\end{tabular}%
	\end{table}

We now discuss the possible reason why directly employing the VPDM in other tasks for SE fails, and then present a customized VPDM as a baseline for our proposed VPIDM.
In SE, which essentially involves reconstructing audio, noisy speech serves as a significant prior. While high-level features like the Mel spectrum \cite{lu_apsipa, lu2022conditional} can be used as input, they often omit crucial low-level features valuable for SE. Therefore, such methods might not yield optimal results. Studies in \cite{welker2022speech,richter_diff} employ low-level features of the raw noisy speech spectrum as input, maintaining the same dimensionality as clean speech for effective SE. We find this strategy also works well for VPDM, namely, the state variable is concatenated on noisy speech and then fed into ANN. For a given state index $\tau$, ANN in VPDM aims to estimate Gaussian noise within
the state variable $\mathbf{S}(\tau)$ \cite{ho2020denoising, song2019generative}. However, in this context,
ANN can predict the clean signal directly from noisy speech and infer Gaussian noise implicitly \cite{kingma2021variational}. Typically, Gaussian noise in $\mathbf{S}(\tau)$ has a higher average energy than the target noise in noisy speech. When processing raw noisy speech, ANN tends to first estimate the clean signal from the noisy one and then derive the Gaussian component. This approach, while simpler, may lessen ANN's ability to learn the entire diffusion process effectively. VPIDM could be considered as applying an interpolating scheme to VPDM.
When there is no target noise or the interpolating coefficient is zero, VPIDM simplifies to VPDM, thus positioning VPDM as a special case within the broader VPIDM framework. As one of our baselines, we implemented a VPDM using the same data set, hyper-parameters, and other experimental configurations as those in VPIDM. The only difference is that we set the interpolating coefficient $1 - \lambda(\tau)\equiv0$ for VPDM.

Compared to VPDM, VPIDM utilizes an interpolation scheme that provides guidance to remove target noise during the reverse process. Although the interpolation approach is adopted in \cite{lu2022conditional, welker2022speech, richter_diff}, none of them provide a theoretical analysis of the mechanism. Inspired by this, we will present an analysis of the mechanism behind our interpolation approach in the next sub-section.


\subsection{Role of Interpolation in Enhancing Speech}
\label{reducing_noise_reason}
In the reverse process, the objective is to sample a clean speech signal starting from the initial state $\mathbf{S}(T)$. This process reverses the deterministic process, denoted as $\mathbf{U}(\tau)$, and the stochastic process.
The stochastic process involves reducing all the Gaussian noise present in $\mathbf{S}(T)$. Typically, this sampling is carried out using R-SODE in Eq. (\ref{idm_re_nn}), with the drift coefficient $\mathbf{f}(\mathbf{S}, \mathbf{Y}, \tau)$ specified in Eq. (\ref{drift_co_idm}).
Simultaneously, the deterministic process incrementally removes the target noise via R-SODE. However, understanding the underlying mechanism of how VPIDM effectively eliminates the target noise can be challenging.
By referring to the RTNP defined in Eq. (\ref{add_noise_n}), we can derive the drift function in another form, which turns out to be equivalent to the drift coefficient described in Eq. (\ref{drift_co_idm}),  representing the function for noise as:
\begin{align}
    \mathbf{f}(\mathbf{S}, \mathbf{N}, \tau) &= \frac{\text{d}\ln{\alpha(\tau)}}{\text{d}\tau}\mathbf{S}(\tau) + \alpha(\tau)\frac{\text{d}\eta(\tau)}{\text{d}\tau} \mathbf{N}.
    \label{drift_co_idm_n}
\end{align}
Although the target noise $\mathbf{N}$ is not practically \textcolor{black}{accessible}, the drift function $\mathbf{f}(\mathbf{S}, \mathbf{N}, \tau)$
provides valuable insights into the mechanism. Combine Eqs. (\ref{drift_co_idm_n}), (\ref{mean}) and (\ref{idm_re_nn}), we get:
\newcommand*\tcircle[1]{%
    \textcircled{\fontsize{7pt}{0}\fontfamily{phv}\selectfont #1}%
  \raisebox{-0.5pt}{%
  }%
}
    \begin{align}
        \frac{\text{d}\mathbf{U}(\tau)}{\text{d}\tau} = \underbrace{\vphantom{\frac{\text{d}{\alpha(\tau)}}{\text{d}\tau}\frac{\mathbf{U}(\tau)}{\alpha(\tau)}}\frac{\text{d}{\alpha(\tau)}}{\text{d}\tau}\frac{\mathbf{U}(\tau)}{\alpha(\tau)}}_{\tcircle{1}} + \underbrace{\vphantom{\frac{\text{d}{\alpha(\tau)}}{\text{d}\tau}\frac{\mathbf{U}(\tau)}{\alpha(\tau)}}\alpha(\tau)\frac{\text{d}\eta(\tau)}{\text{d}\tau} \mathbf{N}}_{\tcircle{2}},
    \end{align}
where the first item \tcircle{1} contributes to \textcolor{black}{iteratively building up the complex value, consequently enhancing the clean signal amplitude incrementally over $\tau$ from $T$ to $0$}. The second item \tcircle{2} is to decrease the target noise component. Therefore, noisy speech $\mathbf{Y}$ in the drift coefficient $\mathbf{f}(\mathbf{S}, \mathbf{Y}, \tau)$ has two main contributions, i.e., the clean components in noisy speech help repair the amplitude of estimated clean speech, and the noise part offsets the target noise gradually.

For VEIDM, $\alpha(\tau)=1$, the first item \tcircle{1} is zero. So, noisy speech has only one contribution during the reverse process: providing the noise component to reduce the target noise.

\subsection{VPIDM as a Frontend for Recognizing Noisy Speech}
\label{subsec_asr}

Studies \cite{tu2020multi, gao2016snr, sun2017multiple} highlight that ASR systems exhibit varying levels of sensitivity to noise compared to perceptual metrics. Furthermore, research \cite{IwamotoODISAK22} indicates that ASR systems are more susceptible to artificial interferences than to noise. This sensitivity presents a challenge for SE algorithms: while they can effectively remove target noise, they often introduce artifacts that are detrimental to ASR applications. This situation creates a trade-off between the intensity of noise reduction and the generation of artifacts. Intense denoising tends to produce more artifacts, which can adversely affect recognition performance, and vice versa.

In our model, during the reverse process, an estimated clean speech $\mathbf{S}(\tau)$ is derived from the output of an ANN. This output not only gradually eliminates the Gaussian noise through the R-SODE in Eq. (\ref{idm_re_nn}) but also provides an implicit estimation of $\mathbf{V}(\tau)$. As $\tau$ decreases from $T$ to $0$, the target noise in $\mathbf{V}(\tau)$ is progressively reduced. Both $\mathbf{S}(\tau)$ and $\mathbf{V}(\tau)$ offer mid-outputs with varying noise reduction intensities. When $\tau \rightarrow 0$, the target noise is almost entirely removed, but this often results in distortion of the clean speech components, a.k.a, artifacts.

Previous studies \cite{tu2020multi, gao2016snr, sun2017multiple} suggest that retaining some noise can be beneficial for ASR systems, helping to avoid artifacts. Therefore, we propose using the mid-outputs of ${\mathbf{S}}(\tau)$ and ${\mathbf{V}}(\tau)$ for ASR. While $\mathbf{S}(\tau)$ also provides mid-outputs with moderate denoising intensity, it contains more Gaussian noise than $\mathbf{V}(\tau)$, especially when $\tau$ is close to $T$. Consequently, for noise-robust ASR, we prefer the mid-outputs of $\mathbf{V}(\tau)$, which require fewer steps than the complete reverse process. We determine the optimal number of sampling steps, $K_1 (\le K)$, for obtaining these mid-outputs based on ASR performance on a development dataset, where $K$ is the total number of steps in the discrete reverse process. \textcolor{black}{In other words, during the recursive sampling process, we utilize the estimated $\hat{\mathbf{V}}_{K-K_1}$ taking $K_1$ steps for ASR, and $\hat{\mathbf{S}}_{0}$ taking $K$ steps for SE.}

\section{Experiments and Result Analysis}
\subsection{Experimental Settings}

\subsubsection{Speech Data Sets}
Our models are trained on two well-known datasets: the smaller Voice Bank + Demand (VBD) dataset \cite{valentini2016investigating}, and the larger-scale dataset from the third Deep Noise Suppression Challenge (DNS) \cite{reddy2020interspeech}, to demonstrate our methodology. For selecting the optimal model in the VPIDM training, we randomly chose 20 clips from the validation dataset. When training the discriminative model, all clips in the validation dataset are used for validation to prevent overfitting.

\begin{itemize}
\item{The VBD Corpus}: VBD \cite{valentini2016investigating} is widely adopted for SE tasks. The training set consists of $28$ speakers ($14$ female and $14$ male) with $8$ noises in five SNR levels. The total duration is about $9$ hours. The test set includes two unseen speakers with two unseen noises at five unseen SNR levels. Different from our previous paper \cite{guo23_interspeech} where we use partial clips from the test set for validating,  we preserve clips of two speakers as a validation set, and clips from other $26$ speakers for training.
\item{The DNS Corpus}: The clean DNS set \cite{reddy2020interspeech} consists of $7$ kinds of languages, i.e., English, Mandarin, German, French, Italian, Russian, Spanish, and two kinds of unusual speech, i.e, singing voice, and emotional speech. The total duration is about $660$ hours. About $60000$ noises are provided for simulation. We keep 200 kinds of noises for making up the validation dataset. The 200 clips of clean speech are randomly selected from the unseen TIMIT \footnote{\url{https://catalog.ldc.upenn.edu/LDC93s1}} dataset for validation.  In this study, we discuss the problem based on additive noise. Therefore, the blind dataset without the reverberation in the simulation datasets provided by the challenge organizer is adopted as the test set, denoted as ``DNS Simu''.

\item{The 4th CHiME test data set (CHiME-4)}: The CHiME-$4$ \cite{vincent_2016} test set includes two subsets, i.e., the simulated one and the real one. Each subset consists of $1320$ noisy clips recorded in four real noisy environments, i.e., bus, cafeteria, pedestrian, and street, which is detrimental for the ASR. The real subset means that all clips are recorded from the real noisy environments. To validate VPIDM for the noise-robust ASR, we use the simulated subset as the validation set to select the $K_1$ and test the VPIDM on the real subset trained on the large-scale DNS dataset. It is worth pointing out that this data set is quite different from our DNS training set. For example, both speakers and noises in CHiME-$4$ are unseen for the trained model, besides, we use the simulated data to get clean-noisy pairs for training which have different distributions from the real recorded data.
\end{itemize}

\subsubsection{Neural Networks and Training Settings}
\label{neural_network}
In studies \cite{song2020score}, authors propose a UNet-like ANN architecture, a.k.a, \textit{Noise Conditional Score Network ++} (NCSN++) for image generation. Literature \cite{richter_diff} modified it for the SE task. In this study,  we utilize all the same ANN to validate our proposed method fairly. More detailed settings can be found in \cite{song2020score, richter_diff}. The two hyper-parameters, $\beta_{\text{max}}$ and $\beta_{\text{min}}$, defining the scale coefficient $\alpha(\tau)$ in Eq. (\ref{vpidm_alpha}) are set to $2$ and $0.1$, respectively, we set the $\gamma$ in the $\lambda(\tau)$ to $1.5$, and $T=1$.  The two constants, $a$ and $c$ in the signal model in Eq. (\ref{compress_function}), are set to $0.15$ and $0.5$, respectively. The \textcolor{black}{Hann} window is selected in STFT, with a hop size of $128$ and a window length of $510$. All clips are cut or padded to $256$ frames, resulting in $L=M=256$. We set the effective batch size $4\times8=32$, learning rate  $10^{-4}$.  We train the models for $120$ and $200$ epochs for the VBD and DNS data sets, respectively.
\begin{table*}[th]
    \caption{The overall performance comparison of the VPIDM and four baselines on the VBD dataset.
     Models denoted with asterisks (*) indicate that we have reproduced their results from the respective articles. We have replicated the remaining models based on the settings specified in the corresponding papers. The ``G'' denotes the generative model, and the ``D'' represents the discriminative model. All experimental results are presented in the form of mean $\pm$ standard deviation.}
    \label{tab:trainedon_vbd}
    \centering
    \begin{tabular}{ lccccccc}
        \toprule
        {Methods} &
        Type &
        {PESQ $\uparrow$} &
        {ESTOI (\%) $\uparrow$} &
        CSIG $\uparrow$ &
        CBAK $\uparrow$ &
        COVL $\uparrow$
         \\
        \midrule
         Noisy& -&    $1.97\pm0.75$   &${78.67\pm14.94}$  &$3.35\pm0.87$ &$2.44\pm0.67$ &$2.63\pm0.83$ \\
          \rowcolor[gray]{0.9}MP-SENet$^{*}$ \cite{lu2023mp}& D&   $\mathbf{3.49}\pm \mathbf{0.61}$   &$\mathbf{89.11}\pm\mathbf{8.39}$ &$\mathbf{4.64} \pm \mathbf{0.72}$  &$\mathbf{3.72}\pm\mathbf{0.43}$  &$\mathbf{4.12} \pm \mathbf{0.59}$ \\
       \rowcolor[gray]{0.9} MetricGAN+$^{*}$ \cite{fu21_interspeech}& D&   $3.13\pm0.55$   &${83.15\pm11.20}$ &$4.10\pm0.68$ &$2.89\pm0.40$ &$3.60\pm0.64$ \\
        \rowcolor[gray]{0.9}NCSN++ \cite{song2019generative}& D&   $2.87\pm0.74$   &${87.26\pm9.88}$ &$3.67\pm0.97$ &$3.45\pm0.61$ &$3.25\pm0.88$ \\
        $\text{VEIDM}$ \cite{richter_diff}& G&   $2.93\pm0.63$   &${86.36\pm9.82}$ &$4.12 \pm 0.68$ &$3.37\pm0.36$   &$3.51\pm 0.67$ \\
         $\text{VPIDM}$ (Ours) & G&  ${3.16}\pm {0.69}$   &${87.44}\pm{9.44}$ &${4.23} \pm {0.66}$  &${3.53}\pm{0.53}$  &${3.70} \pm {0.71}$ \\

        \bottomrule
		\end{tabular} %
	\end{table*}

\subsubsection{Evaluation Metrics}
\begin{itemize}
\item{CSIG, CBAK, COVL}:  Signal quality (CSIG), background noise (CBAK), and the overall mean opinion score (COVL) \footnote{\url{https://github.com/mkurop/composite-measure}} in \cite{hu2007evaluation} are adopted to assess the speech quality, extent of reducing noise, and the overall speech quality compared to clean speech. All scales are in $[0, 5]$.
\item{ESTOI and PESQ}:  Extended Short-Time Objective Intelligibility (ESTOI) \footnote{\url{https://github.com/mpariente/pystoi}} \cite{jensen2016algorithm} scaled in $[0, 1]$ and wide-band perceptual evaluation of speech quality (PESQ)\footnote{\url{https://github.com/ludlows/PESQ}} scaled in $[1, 4.5]$ are adopted for evaluate the speech quality and intelligibility. We re-express ESTOI in the percentage form and always use wide-band PESQ here.
\item{WER and ASR backend}: We use word error rate (WER) to measure VPIDM's performance for ASR of noisy speech. \textcolor{black}{ Although there are more sophisticated ASR systems with possibly higher accuracy, it should be noted that the primary objective of our paper is not to showcase the best performance of ASR systems but rather to highlight the relative improvements of early-stopping strategy compared to the final result during the reverse process. Therefore, we follow studies \cite{Chen2018BuildingSD, nianpl, IwamotoODISAK22}, using the light-weighted ASR model \footnote{\url{https://github.com/kaldi-asr/kaldi/tree/master/egs/chime4/s5_1ch}} to unveil the relative improvement.}  The model uses a time delay neural network (TDNN) based on the lattice-free version of maximum mutual information (LF-MMI) trained on all training clips with data augmentation. The language model is a $5$-gram recurrent NN-based language model (RNNLM).
\end{itemize}
\begin{figure}[tb]
  \centering
  \centering{\includegraphics[width=7.5cm]{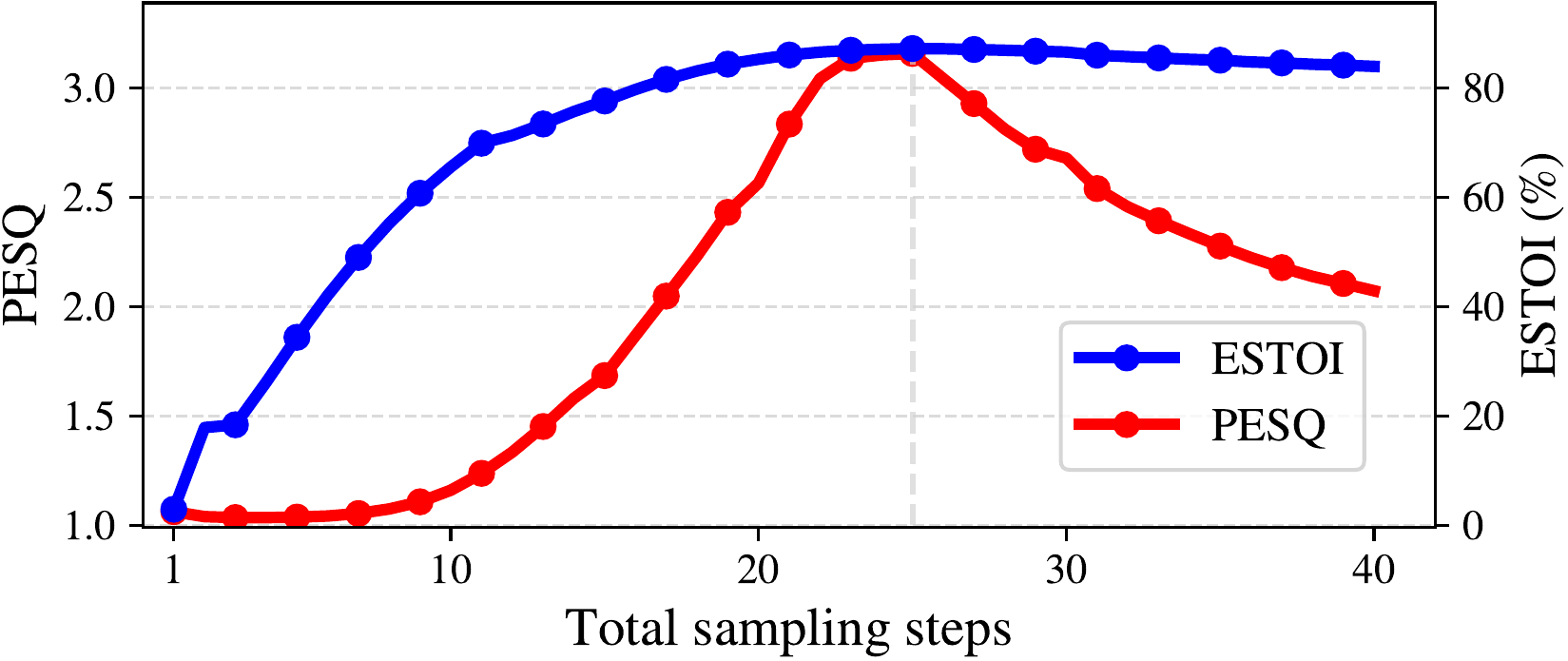}}
%
\caption{The changing trends of two evaluation metrics PESQ/ESTOI concerning the total sampling steps. The number of optimal sampling steps is $25$.}
\label{fig:estoi_steps}
\end{figure}

\subsection{Speech Enhancement Results on VBD}
The results on the VBD set are presented in Table \ref{tab:trainedon_vbd}. Different from our previous paper \cite{guo23_interspeech} in which we adopt partial speech segments from the test set for validation when trained on VBD, here we randomly choose 20 clips from the validation set.
Although the results are slightly different, the same conclusion in \cite{guo23_interspeech} can be drawn. We already demonstrate that the optimal $\epsilon$ is $4\cdot10^{-2}$ \cite{guo23_interspeech}, so we still use this value in this article.

The number of sampling steps in which the best performance is achieved is denoted as the optimal sampling steps (O.S.S.).  To select the O.S.S., we investigate two metrics on the validation set, i.e., PESQ and ESTOI. In Fig. (\ref{fig:estoi_steps}), we give the results of the two metrics changing with the total sampling steps. We can see that both metrics first improve and then deteriorate as the number of sampling steps increases,  and that the best result is achieved when the O.S.S. is $25$.
In addition, the two metric curves increase when the number is less than $25$, this is because when there are too few sampling steps,
the discrete sampling algorithm in Eq. (\ref{reverse_recursive_equation_discrete})
leads to residual Gaussian noise in the enhanced noise.
In other words, fewer steps result in more residual noise.  Moreover, the two metrics curves decrease when the number of sampling steps is greater than $25$,   the reason is that when there are too many steps, the discrete sampling algorithm in Eq. (\ref{reverse_recursive_equation_discrete}) could cause prediction errors which uses the estimated state variable $\mathbf{S}_k$ as one of input while using the ground-truth during the training stage. The prediction errors will be accumulated with the sampling steps and turn into the accumulation of errors. Consequently,  more steps result in more accumulative errors.
Besides, the falling speed is slower than the rising speed. We speculate that the accumulative error is more minor than the residual Gaussian noise.

 Furthermore, we compare the proposed VPIDM to several methods in Table \ref{tab:trainedon_vbd} on the VBD.
\textcolor{black}{The results from discriminative models are shown with a gray background in Table \ref{tab:trainedon_vbd} to emphasize that discriminative and generative algorithms belong to distinct categories.}
\begin{table*}[th]
    \caption{The overall performance comparison of the VPIDM and four baselines on the DNS Simu dataset.}
    \label{tab:trainedon_dns}
    \centering
    \begin{tabular}{ lccccccc}
        \toprule
        {Methods} &
        Type &
        {PESQ $\uparrow$} &
        {ESTOI (\%) $\uparrow$} &
        CSIG $\uparrow$ &
        CBAK $\uparrow$ &
        COVL $\uparrow$
         \\
        \midrule
         Noisy& -&   $1.58\pm0.46$   &${80.99\pm12.19}$ &$3.08 \pm 0.74$ &$2.53\pm0.59$ &$2.29\pm0.60$ \\
         \rowcolor[gray]{0.9}NSNet2$^*$ \cite{NSNet} & D&    $2.38\pm0.56$   &${88.21\pm7.67}$ &$3.85\pm0.57$ &$3.19\pm0.51$ &$3.10\pm0.58$\\
         \rowcolor[gray]{0.9}FSubNet$^*$ \cite{FullSubNet}& D&    $2.89\pm0.67$   &${91.96\pm6.79}$ &$4.20\pm0.68$  &$2.94\pm0.64$  &$3.56\pm0.70$  \\
         \rowcolor[gray]{0.9}NCSN++ \cite{song2019generative}& D&    $2.87\pm0.75$   &${94.15\pm7.49}$ &$3.72\pm0.97$ &$3.78\pm0.66$ &$3.31\pm0.88$ \\
        VEIDM \cite{richter_diff}& G&   $2.93\pm0.67$   &${93.63\pm5.83}$  &$4.34\pm0.60$  &$3.66\pm0.67$ &$3.67\pm0.67$ \\
        VPIDM (Ours)& G&    $\mathbf{3.12}\pm\mathbf{0.66}$   &$\mathbf{94.24}\pm\mathbf{5.46}$ &$\mathbf{4.35}\pm\mathbf{0.61}$ &$\mathbf{3.89}\pm\mathbf{0.63}$ &$\mathbf{3.77}\pm\mathbf{0.68}$ \\
        \bottomrule
		\end{tabular} %
	\end{table*}
\begin{figure*}[htb]
  \centering
  \centering{\includegraphics[width=16cm]{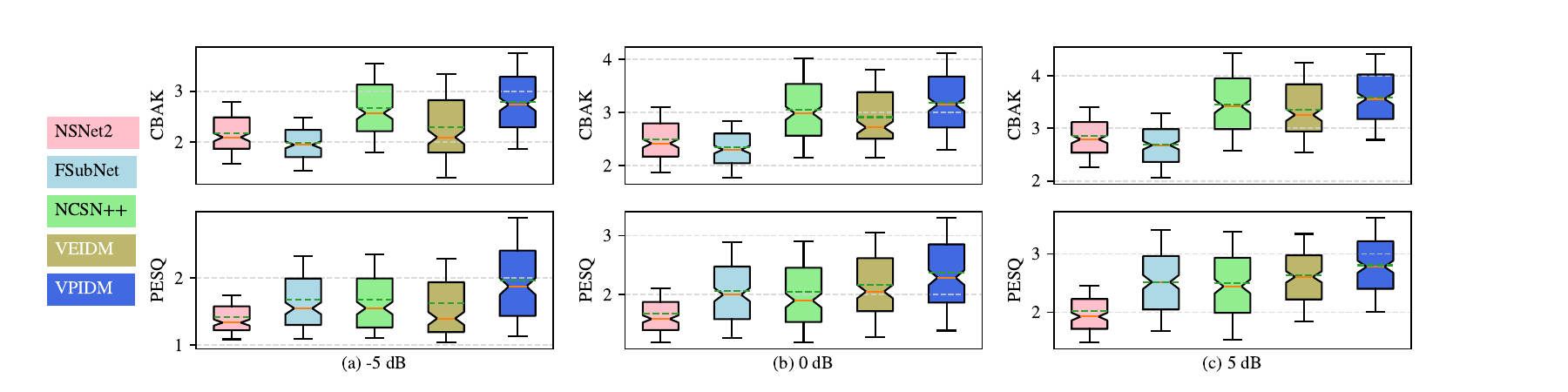}}
%
\caption{Six box plots of two metrics, CBAK and PESQ, for VPIDM and four other baselines at three SNR levels on our simulated data set.  }
\label{fig:performance_on_different_snrs_condition}
\end{figure*}
We already demonstrated that the VPIDM obtains the best results concerning all metrics compared to the current DM-based method in our previous paper \cite{guo23_interspeech}. Thus we only list partial results on the VBD in this article. \textcolor{black}{In addition, as demonstrated in \cite{guo23_interspeech}, when maintaining the same configuration (sample steps and $\epsilon$) as VEIDM, VPIDM still outperforms VEIDM, underscoring the benefits of VP.} Moreover, VEIDM \cite{richter_diff} has already demonstrated its superiority over a series of models. Therefore, in this article,  we only utilize the VEIDM and a new discriminative model i.e., the NCSN++, as our main baselines.
It is worth pointing out that the VEIDM and VPIDM utilize almost the same ANN architecture as that for NCSN++.
C

Compared to the model architecture employed by the VPIDM and VEIDM, the discriminative model NCSN++ means to remove all condition modules related to the state index, parameters of which only account for a small portion of those of the two DMs. In other words, the removed modules almost do not impact the final performance. In our study, we replicated the VEIDM using identical hyper-parameters and configurations described in \cite{richter_diff}. This approach involves utilizing $30$ sampling steps and a corrector, as proposed in \cite{song2020score, richter_diff}, to refine the outcomes during the reverse process. Consequently, this necessitates the model to perform inference twice at each sampling step, leading to about $60$ steps. In this context, the corrector's role is to rectify the estimation errors of the Gaussian component in the state variable during the reverse process of VEIDM.
In contrast, the Gaussian components in the state variables as estimated by our VPIDM at each step are found to be relatively accurate when compared to VEIDM. Therefore, VPIDM eliminates the need for a corrector, streamlining the process and potentially enhancing efficiency.
Although in the shadow of the SOTA discriminative models \cite{cao22_interspeech, lu2023mp},  the proposed VPIDM \textcolor{black}{achieves comparable results and} makes progress towards improving the DM-based method.

\subsection{Speech Enhancement Results on DNS}
Next, we \textcolor{black}{explore} DM-based SE on large-scale data sets \textcolor{black}{which are } not as often studied. \textcolor{black}{We presume that the optimal parameters for models trained on small datasets are also effective on large datasets. Consequently, we train VPIDM and VEIDM using identical model configurations as those used for small-scale datasets, without engaging in hyperparameter selection.} Our experiments are conducted on the DNS set, where we analyze the performance characteristics of our proposed VPIDM in comparison with VEIDM and the discriminative backbone NCSN++.
For baselines, we employ two renowned discriminative models, NSNet2 \cite{NSNet} and FSubNet \cite{ho2020denoising}, trained on the DNS data set. NCSN++ serves as an additional discriminative baseline in our study.

Our results shown in Table \ref{tab:trainedon_dns} demonstrate that both VEIDM and VPIDM outperform NCSN++ in three pivotal metrics: PESQ, CSIG, and COVL, suggesting their efficacy in reconstructing high-quality clean speech. In particular, VPIDM exhibits slightly superior denoising capabilities over NCSN++ in the CBAK metric across both data sets, while VEIDM lags slightly behind in the same metric, indicating less effective noise removal under certain conditions. Despite this, VEIDM maintains competitive signal quality as evidenced by the CSIG metric. This could be attributed to interpolation in both VEIDM and VPIDM in generating high-fidelity clean speech estimates, although VEIDM occasionally misidentifies some target noise as part of the speech component.

To assess the performance of VPIDM in low SNR scenarios, we re-simulate the DNS Simu set to generate three subsets with SNRs of $-5$ dB, $0$ dB, and $5$ dB. Take the data with $0$ dB SNR for example, we use the original clean-noisy pairs in DNS Simu and then modify the original SNR to $0$ dB. We utilize CBAK and PESQ to evaluate the residual background noise and speech quality and present the results in  Fig. \ref{fig:performance_on_different_snrs_condition}.
\begin{figure}[tb]
  \centering
  \centering{\includegraphics[width=8.5cm]{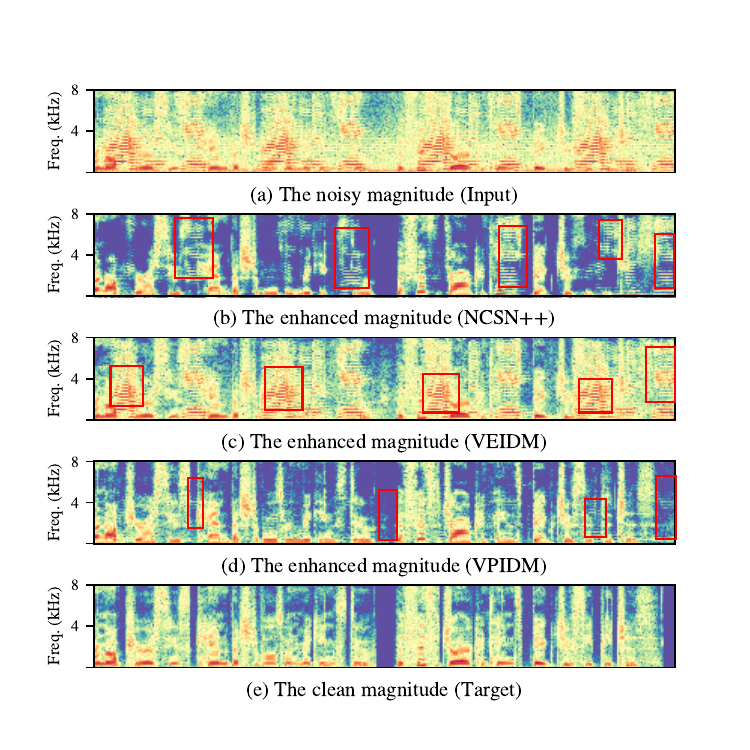}}
%
\caption{Visualization of five magnitude \textcolor{black}{spectra}, displayed in a logarithmic scale, for clean speech, noisy speech, and three speech clips enhanced using NCSN++, VEIDM, and VPIDM, respectively. }
\label{fig:spectrum_on_different_snrs_condition_factory_1}
\end{figure}
Our evaluations reveal that at an input SNR of $-5$ dB, VPIDM attains superior speech quality and lesser residual noise when compared to VEIDM and NCSN++. At this SNR level, VEIDM retains a larger amount of residual noise than that for NCSN++, adversely affecting its PESQ score. At SNRs of $0$ dB and $5$ dB, although VEIDM does not fully eliminate noise like in NCSN++, it achieves better speech quality.
In summary, our proposed VPIDM algorithm outperforms the baseline models in terms of both background noise reduction and speech quality, underscoring its robustness, especially in low SNR conditions.

We illustrate the robustness of our models by drawing spectrograms at $-5$ dB in Fig.~\ref{fig:spectrum_on_different_snrs_condition_factory_1}. The utterance of clean speech is ``clnsp51'' with ``baby cry'' noise. The red rectangles in the figure denote the residual target-noise components introduced by the respective models.
From Fig.~\ref{fig:spectrum_on_different_snrs_condition_factory_1}(b), it is evident that NCSN++ almost completely reduces all target noise but also removes some speech components, a phenomenon known as over-suppression \cite{IwamotoODISAK22}.  In contrast, VEIDM (Fig.~\ref{fig:spectrum_on_different_snrs_condition_factory_1}(c)) only partially removes target noise, retaining many noise components in enhanced speech. VPIDM (Fig.~\ref{fig:spectrum_on_different_snrs_condition_factory_1}(d)), however, not only reduces the target noise but also preserves a significant amount of speech detail, making it closest to clean speech (Fig.~\ref{fig:spectrum_on_different_snrs_condition_factory_1}(e)) among the techniques tested.
\begin{table}[t]

    \caption{A Comparison of  VPIDM and VPDM over four metrics, PESQ, ESTOI, CBAK, and COVL, on the DNS Simu data set.}
    \label{tab_vpidm_vpdm}
    \centering
    \resizebox{0.85\columnwidth}{!}{%
    \begin{tabular}{ lcccc}
        \toprule
        {Settings} &
        {PESQ $\uparrow$} &
        {ESTOI (\%) $\uparrow$} &
         {CBAK  $\uparrow$} &
        {COVL  $\uparrow$}
         \\
        \midrule
        \midrule
        VPDM & $2.68$ &  $91.10$   &${2.68}$   &${3.35}$  \\
        VEIDM & $2.93$ &  $93.63$   &${3.66}$   &${3.67}$ \\
        VPIDM &$3.12$   &$94.24$ & $3.89$    &${3.77}$  \\

        \bottomrule			
		\end{tabular} %
  }
	\end{table}
Furthermore, VEIDM is reported to occasionally produce vocalizing artifacts devoid of linguistic meanings \cite{richter_diff, storm_diff}. In our experiments, VPIDM effectively mitigates this issue, even in low SNR conditions.
Since VEIDM and VPIDM utilize different interpolating schemes, we believe that improving the scheme could further alleviate these artifact noise issues.

Conducting a listening test on the mid-outputs of both VEIDM and VPIDM, we find that the artifact noise problems arise when the estimated state is close to clean speech.
Therefore, our future research direction will involve exploring innovative modifications of the interpolating scheme, coupled with the introduction of advanced techniques to enhance speech quality in state variables as they converge to clean speech. Currently, the network architectures yielding promising results in SE tasks are predominantly adapted from those developed for image generation. However, these models may not fully exploit the characteristics of SE, leaving ample room for improving the model architecture for SE.
Recognizing this trend, our future work will concentrate on investigating and incorporating more advanced network structures specifically tailored to speech enhancement applications.
 \begin{figure}[tb]
  \centering
  \centering{\includegraphics[width=7.5cm]{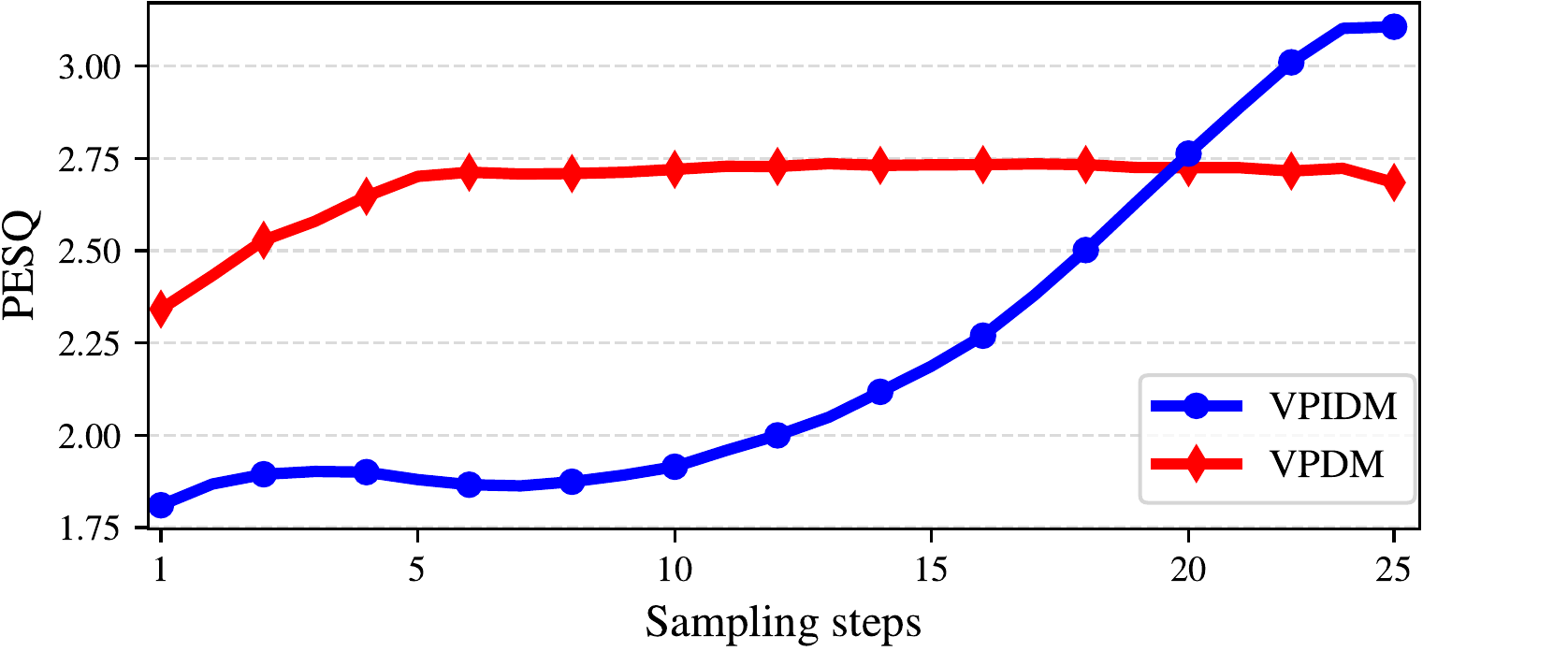}}
%
\caption{The two PESQ curves are used to demonstrate the changes in speech quality during the reverse processes of the VPIDM and VPDM.}
\label{fig:comparison_vpidm_vpdm}
\end{figure}

\subsection{Analysis and Discussion}
\subsubsection{VPDM Versus VPIDM} A comparison of VPDM and VPIDM is presented in Table \ref{tab_vpidm_vpdm}. We observe that VPIDM always outperforms VPDM, which implies the interpolating operation provides better guidance for the ANN to learn the mapping from the initial state to the clean signal during the reverse process. To further demonstrate the impacts, we illustrate the PESQ curves of the deterministic means of VPIDM and VPDM in Fig. \ref{fig:comparison_vpidm_vpdm}. VPDM could predict clean speech (implicitly) with only limited performance and keep the performance almost unchanged after a few sampling steps.
VPIDM, on the other hand, exhibits a gradual performance improvement as the number of sample steps increases.
\begin{figure}[tb]
  \centering
  \centering{\includegraphics[width=7.4cm]{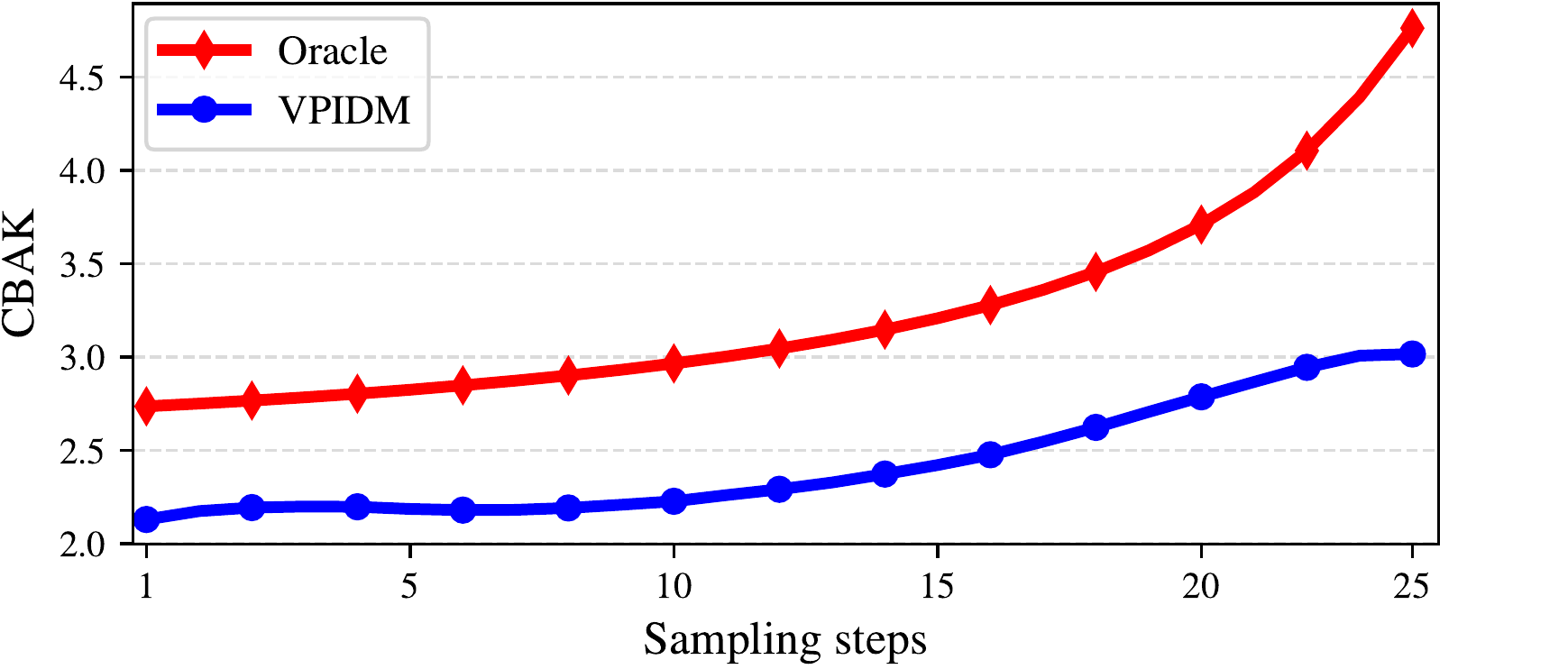}}
%
\caption{The two CBAK curves illustrate variations in denoising intensity for the RTNPs of the VPIDM and the ground truth (denoted as ``Oracle'').}
\label{fig:background_noise}
\end{figure}

\subsubsection{Role of Interpolation in Reducing Target Noise} In Section \ref{reducing_noise_reason}, we present that the target noise is removed gradually by infusing the noisy speech into the mean $\mathbf{U}(\tau)$ during the reverse process, where the mean could be obtained by scaling the RTNP $\mathbf{V}(\tau)$. Therefore, we investigate the $\mathbf{V}(\tau)$ to see how the intensity of the target noise changes.  We draw the CBAK of the ground truth $\mathbf{V}(\tau)$ (denoted as Oracle) and the estimated $\hat{\mathbf{V}}(\tau)$ from the model trained on the DNS dataset to illustrate the process of removing target noise in Fig. \ref{fig:background_noise}.
As the sampling time changes, the noise level in the RTNP is not always monotonically removed like in the Oracle. For example, when the sampling steps are between $4$ and $8$, there is a slight decrease in the curve of the VPIDM, indicating a slight increase in the estimated noise level in $\hat{\mathbf{V}}(\tau)$.
This is because, during the entire reverse process, the estimated Gaussian components of the VPIDM are not entirely accurate at each step,
\begin{figure}[tb]
  \centering
  \centering{\includegraphics[width=7.4cm]{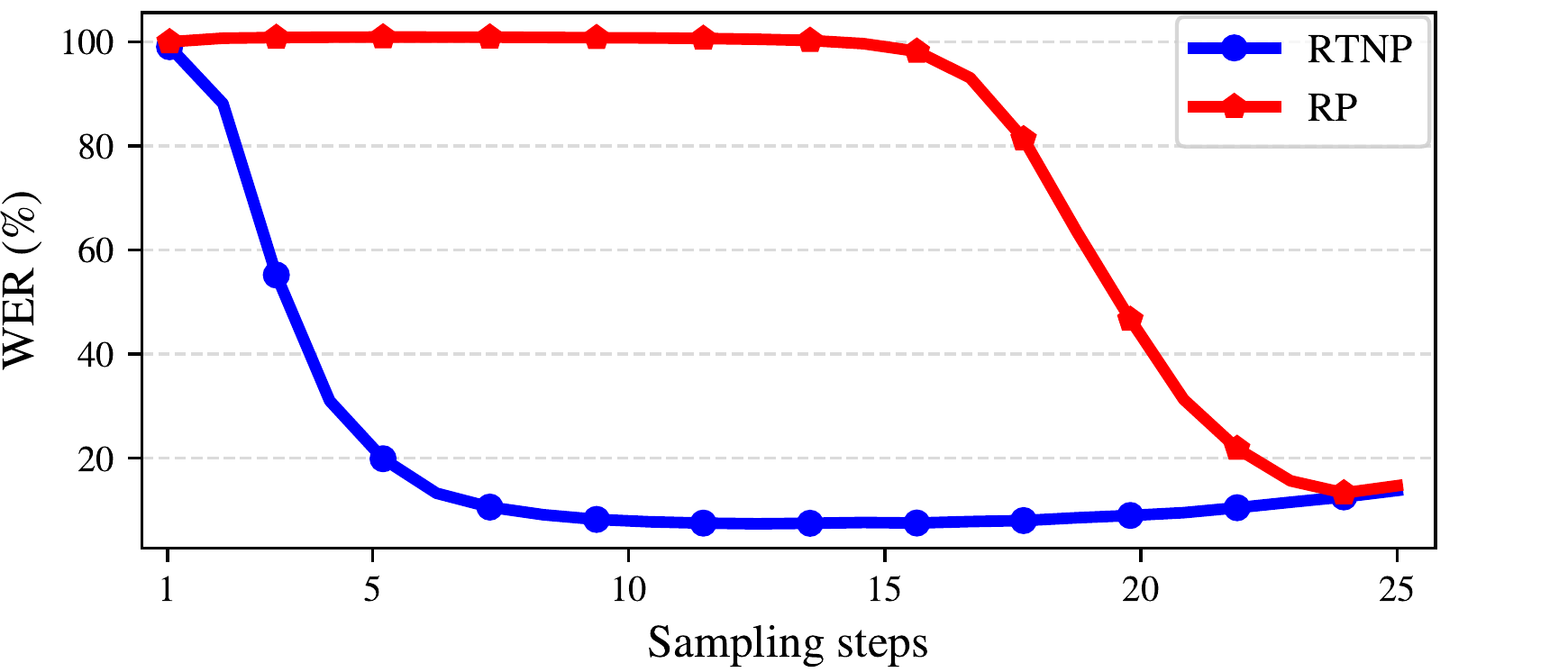}}
%
\caption{The WERs of the mid-outputs of the RTNP (denoted as ``RTNP'') and the reverse process (a.k.a  ``RP'') during the sampling stage. }
\label{fig:vpidm_wer_on_chime4}
\end{figure}
which can result in residual Gaussian noise in the estimated $\mathbf{V}(\tau)$.
The residual Gaussian noise causes slight fluctuations in the local CBAK curve of the RTNP but does not change the global trend. Therefore,
from Fig. \ref{fig:background_noise}, the target noise progressively reduces during the reverse process.

\subsection{ASR Results of Enhanced Speech on CHiME-4} 
\begin{figure}[t]
  \centering
  \centering{\includegraphics[width=7.6cm]{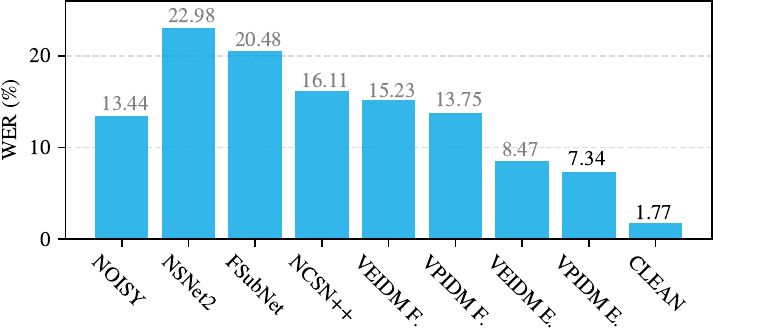}}
%
\caption{A Comparison of WERs with several baselines. ``F'' in VPIDM F and VEIDM F means the final output of the corresponding DM. ``E'' in VPIDM E and VEIDM E denotes the mid-output of the RTNP.}
\label{fig:wer_comparison}
\end{figure}

To further evaluate the generalization capability of VPIDM, which has been trained on the large-scale data set  \textcolor{black}{(DNS corpus)}, we perform tests on the trained models using the CHiME-4 test set for ASR. To demonstrate that the mid-output $\mathbf{V}(\tau)$ of RTNP is more conducive for ASR than the mid-output $\mathbf{S}(\tau)$ of the reverse process, we present the WER performance curves of these two methods in Fig.~\ref{fig:vpidm_wer_on_chime4} on the CHiME-$4$ simulated test set. It is observed that the mid-outputs necessitate a greater number of sampling steps to achieve a competitive WER.
This is attributed to the fact that $\mathbf{S}(\tau)$ is noisier than $\mathbf{V}(\tau)$. Specifically, $\mathbf{S}(\tau)$ encompasses not only the target noise but also the Gaussian noise components, while $\mathbf{V}(\tau)$ contains solely the target noise.
In addition, the speech components in $\mathbf{S}(\tau)$ are scaled down by $\alpha(\tau)$, resulting in a more detrimental condition.
In a similar vein, we implement this strategy for VEIDM trained on the DNS data set.
Fig.~\ref{fig:wer_comparison} presents the WERs for VEIDM (labeled as ``VEIDM E'') and VPIDM (``VPIDM E''),
alongside the WERs for the final outputs of VEIDM and VPIDM (``VEIDM F'' and ``VPIDM F''),
and the WERs from various baseline models. When adopting the mid-output of $\hat{\mathbf{V}}(\tau)$, as depicted in Fig.~\ref{fig:vpidm_wer_on_chime4}, optimal performance is observed around $K/2$. \textcolor{black}{We further conduct an ablation study to select the best $K_1$ on the same development dataset for VEIDM which is trained on the DNS dataset. The best $K_1$ for VEIDM is also approximately $K/2$.}
Practically, we set $K_1$ to approximately $12$ for VPIDM (with a total of $25$ sampling steps) and $14$ for the VEIDM (comprising $30$ sampling steps). Notably, the final outputs of the two DMs demonstrate superior WERs relative to NCSN++, yet they do not outperform raw noisy speech. This suggests that while the two DMs can reconstruct cleaner speech than the discriminative model, they may still compromise speech naturalness, adversely affecting the ASR performances. Fig. \ref{fig:wer_comparison} also indicates that the two DMs outperform all baselines and noisy speech in terms of WERs, highlighting the enhanced efficacy of the two DMs.
\section{Conclusion}
Building upon our prior work in \cite{guo23_interspeech}, this study further develops a new interpolating scheme within the DM framework for single-channel SE. We perform rigorous theoretical derivations and conduct extensive experimental validations of the proposed VPIDM.
We demonstrate that VPIDM is suitable for SE when compared to the VPDM and obtains superior performances over VEIDM when evaluated in both small and large-scale data sets. It is particularly noteworthy to mention VPIDM's robustness in low SNR conditions, where it effectively eliminates target noise and reconstructs clean speech. VPIDM also alleviates issues, such as artificial noises, as mentioned in \cite{richter_diff}, and mitigates the problem of over-suppression of noise in SE.
As an ASR frontend, VPIDM generates estimated clean speech with enhanced spectral detail and demonstrated
effectiveness for robust ASR of noisy speech. However, despite its improved sampling efficiency over VEIDM, VPIDM's computational cost remains high. Future research will focus on reducing the number of sampling steps in DMs to enhance speech more efficiently.
Another future work is to tailor the interpolating schemes to specific application scenarios, such as reverberation, from both theoretical and experimental perspectives.
{\appendix[Derivation of Drift and Diffusion Coefficients]
\label{appendix_derivation}
Consider the drift  $\mathbf{f}(\mathbf{S}, \mathbf{Y}, \tau)$ is an affine function of $\mathbf{S}(\tau)$, according to Eq. (5-50) in \cite{solin_2019}, we get the drift coefficient:
\begin{align}
\label{derivation_of_drift}
    \frac{\text{d}\mathbf{U}}{\text{d}\tau} &= \mathbb{E}_\mathbf{S}\left[\mathbf{f}(\mathbf{S}, \mathbf{Y}, \tau)\right] = \mathbf{f}(\mathbb{E}_\mathbf{S}[\mathbf{S}], \mathbf{Y}, \tau) = \mathbf{f}(\mathbf{U}, \mathbf{Y}, \tau) \nonumber \\
    &= \frac{\text{d}\left[\alpha(\tau)\lambda(\tau)\right]}{\text{d}\tau}\mathbf{X} + \frac{\text{d}\left[\alpha(\tau)(1-\lambda(\tau))\right]}{\text{d}\tau}\mathbf{Y}\nonumber \\
    &=\frac{\text{d}\left[\alpha(\tau)\lambda(\tau)\right]}{\text{d}\tau}\frac{\mathbf{U}(\tau) - \alpha(\tau)(1 - \lambda(\tau))\mathbf{Y}}{\alpha(\tau)\lambda(\tau)} \nonumber\\
    &+\frac{\text{d}\left[\alpha(\tau)(1-\lambda(\tau))\right]}{\text{d}\tau}\mathbf{Y} \nonumber\\
    &= \frac{\text{d}\left[\alpha(\tau)\lambda(\tau)\right]}{\text{d}\tau}\frac{\mathbf{U}(\tau)}{\alpha(\tau)\lambda(\tau)}
    +\frac{\text{d}\lambda(\tau))}{\text{d}\tau}\frac{\alpha(\tau)}{\lambda(\tau)}\mathbf{Y} \nonumber\\
    &= \frac{\text{d}\ln\left[{\alpha(\tau) \lambda(\tau)}\right]}{\text{d}\tau}\mathbf{U}(\tau) - \alpha(\tau)\frac{\text{d}\ln\lambda(\tau)}{\text{d}\tau}\mathbf{Y}.
\end{align}
Substitute $\mathbf{U}(\tau)$ in Eq. (\ref{derivation_of_drift}) with $\mathbf{S}(\tau)$, we get the drift presented in Eq. (\ref{drift_co_idm}). Furthermore, according to Eq. (5-50) in \cite{solin_2019}, we get the diffusion coefficient \textcolor{black}{shown in Eq.(\ref{diff_co_idm})}:
\begin{align}
\label{derivation_diffusion}
    \frac{\text{d}{\mathbf{\Sigma}(\tau)}}{\text{d}\tau}&= \mathbb{\textcolor{black}{E}}_\mathbf{S}\left[\mathbf{f}(\mathbf{S}, \mathbf{Y}, \tau)(\mathbf{S}-\mathbf{U})^{\text{H}}\right] \nonumber\\ &+ \mathbb{\textcolor{black}{E}}_\mathbf{S}\left[(\mathbf{S}-\mathbf{U})\mathbf{f}^{\text{H}}(\mathbf{S}, \mathbf{Y}, \tau)\right] + g^2(\tau)\mathbf{I} \nonumber\\
    &= 2\frac{\text{d}\ln\left[{\alpha(\tau) \lambda(\tau)}\right]}{\text{d}\tau}\mathbb{\textcolor{black}{E}}_\mathbf{S}\left[(\mathbf{S}-\mathbf{U})(\mathbf{S}-\mathbf{U})^{\text{H}}\right] + g^2(\tau)\mathbf{I} \nonumber\\
    &= 2\frac{\text{d}\ln\left[{\alpha(\tau) \lambda(\tau)}\right]}{\text{d}\tau} \mathbf{\Sigma}(\tau)+ g^2(\tau)\mathbf{I}  \nonumber\\
    &= 2\frac{\text{d}\ln\left[{\alpha(\tau) \lambda(\tau)}\right]}{\text{d}\tau} G^2(\tau)\mathbf{I} + g^2(\tau)\mathbf{I}.
\end{align}

\bibliographystyle{IEEEtran}
\balance
\bibliography{refs}

\begin{thebibliography}{10}
\providecommand{\url}[1]{#1}
\csname url@samestyle\endcsname
\providecommand{\newblock}{\relax}
\providecommand{\bibinfo}[2]{#2}
\providecommand{\BIBentrySTDinterwordspacing}{\spaceskip=0pt\relax}
\providecommand{\BIBentryALTinterwordstretchfactor}{4}
\providecommand{\BIBentryALTinterwordspacing}{\spaceskip=\fontdimen2\font plus
\BIBentryALTinterwordstretchfactor\fontdimen3\font minus
  \fontdimen4\font\relax}
\providecommand{\BIBforeignlanguage}[2]{{%
\expandafter\ifx\csname l@#1\endcsname\relax
\typeout{** WARNING: IEEEtran.bst: No hyphenation pattern has been}%
\typeout{** loaded for the language `#1'. Using the pattern for}%
\typeout{** the default language instead.}%
\else
\language=\csname l@#1\endcsname
\fi
#2}}
\providecommand{\BIBdecl}{\relax}
\BIBdecl

\bibitem{chen2006new}
J.~Chen and Y.~Huang, ``New insights into the noise reduction wiener filter,''
  \emph{IEEE Trans. on audio, speech, and lang. processing}, vol.~14, no.~4,
  pp. 1218--1234, 2006.

\bibitem{yu2016automatic}
D.~Yu and L.~Deng, \emph{Automatic speech recognition}.\hskip 1em plus 0.5em
  minus 0.4em\relax Springer, 2016.

\bibitem{vincent_2016}
E.~Vincent, S.~Watanabe, A.~A. Nugraha, and et~al., ``An analysis of
  environment, microphone and data simulation mismatches in robust speech
  recognition,'' \emph{Comput. Speech Lang.}, vol.~46, pp. 535--557, 2017.

\bibitem{you2021knowledge}
C.~You, N.~Chen, and Y.~Zou, ``Knowledge distillation for improved accuracy in
  spoken question answering,'' in \emph{ICASSP}, 2021, pp. 7793--7797.

\bibitem{you2021self}
------, ``Self-supervised contrastive cross-modality representation learning
  for spoken question answering,'' in \emph{Findings of the Association for
  Computational Linguistics: EMNLP 2021}, 2021, pp. 28--39.

\bibitem{you2022end}
C.~You, N.~Chen, F.~Liu, S.~Ge, X.~Wu, and Y.~Zou, ``End-to-end spoken
  conversational question answering: Task, dataset and model,'' in
  \emph{Findings of the Association for Computational Linguistics: NAACL 2022},
  2022, pp. 1219--1232.

\bibitem{youCZ21}
C.~You, N.~Chen, and Y.~Zou, ``Mrd-net: Multi-modal residual knowledge
  distillation for spoken question answering,'' in \emph{Proceedings of the
  Thirtieth International Joint Conference on Artificial Intelligence,
  {IJCAI}}.\hskip 1em plus 0.5em minus 0.4em\relax ijcai.org, 2021, pp.
  3985--3991.

\bibitem{chenYZ21}
N.~Chen, C.~You, and Y.~Zou, ``Self-supervised dialogue learning for spoken
  conversational question answering,'' in \emph{ISCA Interspeech}.\hskip 1em
  plus 0.5em minus 0.4em\relax {ISCA}, 2021, pp. 231--235.

\bibitem{boll1979suppression}
S.~Boll, ``Suppression of acoustic noise in speech using spectral
  subtraction,'' \emph{IEEE Trans. on acoustics, speech, and signal
  processing}, vol.~27, no.~2, pp. 113--120, 1979.

\bibitem{wang2014training}
Y.~Wang, A.~Narayanan, and D.~Wang, ``On training targets for supervised speech
  separation,'' \emph{IEEE/ACM Trans. Audio, Speech, Lang. Process.}, vol.~22,
  no.~12, pp. 1849--1858, 2014.

\bibitem{li2021domain}
Y.~Li, Y.~Sun, K.~Horoshenkov, and S.~M. Naqvi, ``Domain adaptation and
  autoencoder-based unsupervised speech enhancement,'' \emph{IEEE Trans. on
  Artificial Intelligence}, vol.~3, no.~1, pp. 43--52, 2021.

\bibitem{lin2021unsupervised}
H.-Y. Lin, H.-H. Tseng, X.~Lu, and Y.~Tsao, ``Unsupervised noise adaptive
  speech enhancement by discriminator-constrained optimal transport,''
  \emph{Advances in Neural Information Processing Systems}, vol.~34, pp.
  19\,935--19\,946, 2021.

\bibitem{williamson_complex_2016}
D.~S. Williamson, Y.~Wang, and D.~Wang, ``\BIBforeignlanguage{en}{Complex
  {Ratio} {Masking} for {Monaural} {Speech} {Separation}},''
  \emph{\BIBforeignlanguage{en}{IEEE/ACM Trans. Audio, Speech, Lang.
  Process.}}, vol.~24, no.~3, pp. 483--492, Mar. 2016.

\bibitem{lim_oppen}
J.~Lim and A.~Oppenheim, ``Enhancement and bandwidth compression of noisy
  speech,'' \emph{Proceedings of the IEEE}, vol.~67, no.~12, pp. 1586--1604,
  1979.

\bibitem{xu2014regression}
Y.~Xu, J.~Du, L.-R. Dai, and C.-H. Lee, ``A regression approach to speech
  enhancement based on deep neural networks,'' \emph{IEEE/ACM Trans. Audio,
  Speech, Lang. Process.}, vol.~23, no.~1, pp. 7--19, 2014.

\bibitem{tu2020multi}
Y.-H. Tu, J.~Du, T.~Gao, and C.-H. Lee, ``A multi-target snr-progressive
  learning approach to regression based speech enhancement,'' \emph{IEEE/ACM
  Trans. Audio, Speech, Lang. Process.}, vol.~28, pp. 1608--1619, 2020.

\bibitem{luo2019conv}
Y.~Luo and N.~Mesgarani, ``Conv-tasnet: Surpassing ideal time--frequency
  magnitude masking for speech separation,'' \emph{IEEE/ACM Trans. Audio,
  Speech, Lang. Process.}, vol.~27, no.~8, pp. 1256--1266, 2019.

\bibitem{tan_crn}
K.~Tan and D.~Wang, ``Complex spectral mapping with a convolutional recurrent
  network for monaural speech enhancement,'' in \emph{IEEE Int. Conf. Acoust.,
  Speech, Signal Process.}, 2019, pp. 6865--6869.

\bibitem{cao22_interspeech}
R.~Cao, S.~Abdulatif, and B.~Yang, ``{CMGAN: Conformer-based Metric GAN for
  Speech Enhancement},'' in \emph{ISCA Interspeech}, 2022, pp. 936--940.

\bibitem{chen2015speech}
Z.~Chen, S.~Watanabe, H.~Erdogan, and J.~R. Hershey, ``Speech enhancement and
  recognition using multi-task learning of long short-term memory recurrent
  neural networks,'' in \emph{Sixteenth Annual Conf. of the International
  Speech Communication Association}, 2015.

\bibitem{wu2006two}
M.~Wu and D.~Wang, ``A two-stage algorithm for one-microphone reverberant
  speech enhancement,'' \emph{IEEE Trans. Audio, Speech, Lang. Process.},
  vol.~14, no.~3, pp. 774--784, 2006.

\bibitem{gao2016snr}
T.~Gao, J.~Du, L.-R. Dai, and C.-H. Lee, ``Snr-based progressive learning of
  deep neural network for speech enhancement.'' in \emph{ISCA Interspeech},
  2016, pp. 3713--3717.

\bibitem{bie2022unsupervised}
X.~Bie, S.~Leglaive, X.~Alameda-Pineda, and L.~Girin, ``Unsupervised speech
  enhancement using dynamical variational autoencoders,'' \emph{IEEE/ACM Trans.
  Audio, Speech, Lang. Process.}, vol.~30, pp. 2993--3007, 2022.

\bibitem{strauss2021flow}
M.~Strauss and B.~Edler, ``A flow-based neural network for time domain speech
  enhancement,'' in \emph{IEEE Int. Conf. Acoust., Speech, Signal Process.},
  2021, pp. 5754--5758.

\bibitem{fu2019metricgan}
S.-W. Fu, C.-F. Liao, Y.~Tsao, and S.-D. Lin, ``Metricgan: Generative
  adversarial networks based black-box metric scores optimization for speech
  enhancement,'' in \emph{Int. Conf. on Machine Learning}.\hskip 1em plus 0.5em
  minus 0.4em\relax PMLR, 2019, pp. 2031--2041.

\bibitem{phan2020improving}
H.~Phan, I.~V. McLoughlin, L.~Pham, and et~al., ``Improving gans for speech
  enhancement,'' \emph{IEEE Signal Processing Letters}, vol.~27, pp.
  1700--1704, 2020.

\bibitem{richter_diff}
J.~Richter, S.~Welker, J.-M. Lemercier, and et~al., ``Speech enhancement and
  dereverberation with diffusion-based generative models,'' \emph{IEEE/ACM
  Trans. Audio, Speech, Lang. Process.}, vol.~31, pp. 2351--2364, 2023.

\bibitem{lu_apsipa}
Y.-J. Lu, Y.~Tsao, and S.~Watanabe, ``A study on speech enhancement based on
  diffusion probabilistic model,'' in \emph{Asia-Pacific Signal and Information
  Processing Association Annual Summit and Conf.}, 2021, pp. 659--666.

\bibitem{zhang21c_interspeech}
J.~Zhang, S.~Jayasuriya, and V.~Berisha, ``{Restoring Degraded Speech via a
  Modified Diffusion Model},'' in \emph{ISCA Interspeech}, 2021, pp. 221--225.

\bibitem{lu2022conditional}
Y.-J. Lu, Z.-Q. Wang, S.~Watanabe, A.~Richard, C.~Yu, and Y.~Tsao,
  ``Conditional diffusion probabilistic model for speech enhancement,'' in
  \emph{IEEE Int. Conf. Acoust., Speech, Signal Process.}, 2022, pp.
  7402--7406.

\bibitem{welker2022speech}
S.~Welker, J.~Richter, and T.~Gerkmann, ``Speech enhancement with score-based
  generative models in the complex {STFT} domain,'' in \emph{ISCA Interspeech},
  2022, pp. 2928--2932.

\bibitem{guo23_interspeech}
Z.~Guo, J.~Du, C.-H. Lee, and et~al., ``{Variance-Preserving-Based
  Interpolation Diffusion Models for Speech Enhancement},'' in \emph{ISCA
  Interspeech}, 2023, pp. 1065--1069.

\bibitem{moliner_diff}
E.~Moliner, J.~Lehtinen, and V.~Välimäki, ``Solving audio inverse problems
  with a diffusion model,'' in \emph{IEEE Int. Conf. Acoust., Speech, Signal
  Process.}, 2023.

\bibitem{srtnet}
Z.~Qiu, M.~Fu, Y.~Yu, L.~Yin, F.~Sun, and H.~Huang, ``Srtnet: Time domain
  speech enhancement via stochastic refinement,'' in \emph{IEEE Int. Conf.
  Acoust., Speech, Signal Process.}, 2023.

\bibitem{cold_diff}
H.~Yen, F.~G. Germain, G.~Wichern, and J.~L. Roux, ``Cold diffusion for speech
  enhancement,'' in \emph{IEEE Int. Conf. Acoust., Speech, Signal Process.},
  2023.

\bibitem{storm_diff}
J.-M. Lemercier, J.~Richter, S.~Welker, and T.~Gerkmann, ``Storm: A
  diffusion-based stochastic regeneration model for speech enhancement and
  dereverberation,'' \emph{IEEE/ACM Trans. Audio, Speech, Lang. Process.},
  vol.~31, pp. 2724--2737, 2023.

\bibitem{un_diff}
K.~Saito, N.~Murata, T.~Uesaka, and et~al., ``Unsupervised vocal
  dereverberation with diffusion-based generative models,'' in \emph{IEEE Int.
  Conf. Acoust., Speech, Signal Process.}, 2023.

\bibitem{rl_diff}
C.~Chen, Y.~Hu, W.~Weng, and E.~S. Chng, ``Metric-oriented speech enhancement
  using diffusion probabilistic model,'' in \emph{IEEE Int. Conf. Acoust.,
  Speech, Signal Process.}, 2023.

\bibitem{song2019generative}
Y.~Song and S.~Ermon, ``Generative modeling by estimating gradients of the data
  distribution,'' \emph{Advances in neural information processing systems},
  vol.~32, 2019.

\bibitem{ho2020denoising}
J.~Ho, A.~Jain, and P.~Abbeel, ``Denoising diffusion probabilistic models,''
  \emph{Advances in Neural Information Processing Systems}, vol.~33, pp.
  6840--6851, 2020.

\bibitem{rombach2022high}
R.~Rombach, A.~Blattmann, D.~Lorenz, and et~al., ``High-resolution image
  synthesis with latent diffusion models,'' in \emph{the IEEE/CVF conf. on
  computer vision and pattern recognition}, 2022, pp. 10\,684--10\,695.

\bibitem{diff-wav}
Z.~Kong, W.~Ping, J.~Huang, and et~al., ``Diffwave: {A} versatile diffusion
  model for audio synthesis,'' in \emph{ICLR}, 2021.

\bibitem{popov2021diffusion}
V.~Popov, I.~Vovk, V.~Gogoryan, T.~Sadekova, M.~S. Kudinov, and J.~Wei,
  ``Diffusion-based voice conversion with fast maximum likelihood sampling
  scheme,'' in \emph{ICLR}, 2021.

\bibitem{song2020score}
Y.~Song, J.~Sohl{-}Dickstein, D.~P. Kingma, and et~al., ``Score-based
  generative modeling through stochastic differential equations,'' in
  \emph{ICLR}, 2021.

\bibitem{solin_2019}
S.~Särkkä and A.~Solin, \emph{Applied Stochastic Differential
  Equations}.\hskip 1em plus 0.5em minus 0.4em\relax Cambridge University
  Press, 2019.

\bibitem{kingma2021variational}
D.~Kingma, T.~Salimans, B.~Poole, and J.~Ho, ``Variational diffusion models,''
  \emph{Advances in neural information processing systems}, vol.~34, pp.
  21\,696--21\,707, 2021.

\bibitem{petersen2008matrix}
K.~B. Petersen, M.~S. Pedersen \emph{et~al.}, ``The matrix cookbook,''
  \emph{Technical University of Denmark}, vol.~7, no.~15, p. 510, 2008.

\bibitem{analysis}
B.~Liu, J.~Tao, Z.~Wen, and F.~Mo, ``Speech enhancement based on
  analysis--synthesis framework with improved parameter domain enhancement,''
  \emph{Journal of Signal Processing Systems}, vol.~82, pp. 141--150, 2016.

\bibitem{vincent2011connection}
P.~Vincent, ``A connection between score matching and denoising autoencoders,''
  \emph{Neural computation}, vol.~23, no.~7, pp. 1661--1674, 2011.

\bibitem{sun2017multiple}
L.~Sun, J.~Du, L.-R. Dai, and C.-H. Lee, ``Multiple-target deep learning for
  lstm-rnn based speech enhancement,'' in \emph{Hands-free Speech
  Communications and Microphone Arrays}.\hskip 1em plus 0.5em minus 0.4em\relax
  IEEE, 2017, pp. 136--140.

\bibitem{IwamotoODISAK22}
K.~Iwamoto, T.~Ochiai, M.~Delcroix, and et~al., ``How bad are artifacts?:
  Analyzing the impact of speech enhancement errors on {ASR},'' in \emph{ISCA
  Interspeech}, 2022, pp. 5418--5422.

\bibitem{valentini2016investigating}
C.~Valentini-Botinhao, X.~Wang, S.~Takaki, and J.~Yamagishi, ``Investigating
  rnn-based speech enhancement methods for noise-robust text-to-speech.'' in
  \emph{SSW}, 2016, pp. 146--152.

\bibitem{reddy2020interspeech}
C.~K. Reddy, E.~Beyrami, H.~Dubey, and et~al., ``The interspeech 2020 deep
  noise suppression challenge: Datasets, subjective testing framework, and
  challenge results,'' in \emph{ISCA Interspeech}, 2020.

\bibitem{lu2023mp}
Y.-X. Lu, Y.~Ai, and Z.-H. Ling, ``{MP-SENet}: A speech enhancement model with
  parallel denoising of magnitude and phase spectra,'' in \emph{ISCA
  Interspeech}, 2023, pp. 3834--3838.

\bibitem{fu21_interspeech}
S.-W. Fu, C.~Yu, T.-A. Hsieh, P.~Plantinga, M.~Ravanelli, X.~Lu, and Y.~Tsao,
  ``{MetricGAN+: An Improved Version of MetricGAN for Speech Enhancement},'' in
  \emph{ISCA Interspeech}, 2021, pp. 201--205.

\bibitem{hu2007evaluation}
Y.~Hu and P.~C. Loizou, ``Evaluation of objective quality measures for speech
  enhancement,'' \emph{IEEE Trans. on audio, speech, and lang. processing},
  vol.~16, no.~1, pp. 229--238, 2007.

\bibitem{jensen2016algorithm}
J.~Jensen and C.~H. Taal, ``An algorithm for predicting the intelligibility of
  speech masked by modulated noise maskers,'' \emph{IEEE/ACM Trans. Audio,
  Speech, Lang. Process.}, vol.~24, no.~11, pp. 2009--2022, 2016.

\bibitem{Chen2018BuildingSD}
S.-J. Chen, A.~S. Subramanian, H.~Xu, and S.~Watanabe, ``Building
  state-of-the-art distant speech recognition using the chime-4 challenge with
  a setup of speech enhancement baseline,'' in \emph{ISCA Interspeech}, 2018.

\bibitem{nianpl}
Z.~Nian, J.~Du, Y.~Ting~Yeung, and R.~Wang, ``A time domain progressive
  learning approach with snr constriction for single-channel speech enhancement
  and recognition,'' in \emph{IEEE Int. Conf. Acoust., Speech, Signal
  Process.}, 2022, pp. 6277--6281.

\bibitem{NSNet}
S.~Braun and I.~Tashev, ``Data augmentation and loss normalization for deep
  noise suppression,'' in \emph{Speech and Computer - 22nd Int. Conf., {SPECOM}
  2020}, vol. 12335.\hskip 1em plus 0.5em minus 0.4em\relax Springer, 2020, pp.
  79--86.

\bibitem{FullSubNet}
X.~Hao, X.~Su, R.~Horaud, and X.~Li, ``Fullsubnet: {A} full-band and sub-band
  fusion model for real-time single-channel speech enhancement,'' in \emph{IEEE
  Int. Conf. Acoust., Speech, Signal Process.}\hskip 1em plus 0.5em minus
  0.4em\relax {IEEE}, 2021, pp. 6633--6637.

\end{thebibliography}

\end{document}


\begin{table*}[th]
    
    \caption{The.}
    \label{tab:general}
    \centering
    \begin{tabular}{ lcccccccc}
        \toprule
        {Methods} & 
        Type &
        Dataset &
       
        {PESQ $\uparrow$} &
        {ESTOI (\%) $\uparrow$} &
        CSIG $\uparrow$ &
        CBAK $\uparrow$ &
        COVL $\uparrow$ &
        {SI-SDR $\uparrow$}    
         \\
        \midrule  
        \midrule
        Noisy& -& VBD.&   $1.97\pm0.75$   &${78.67\pm14.94}$  &$3.35\pm0.87$ &$2.44\pm0.67$ &$2.63\pm0.83$ &$8.45\pm5.62$ \\     
        MetricGAN+ \cite{fu21_interspeech}& D& VBD.&   $3.13\pm0.55$   &${83.15\pm11.20}$ &$4.10\pm0.68$ &$2.89\pm0.40$ &$3.60\pm0.64$ &$8.53\pm3.83$ \\ 
        NCSN++ \cite{song2019generative}& D& VBD.&  $2.87\pm0.74$   &${87.26\pm9.88}$ &$3.67\pm0.97$ &$3.45\pm0.61$ &$3.25\pm0.88$ &$\mathbf{19.74}\pm\mathbf{3.63}$\\ 
        $\text{VEIDM}_\text{n}$ \cite{richter_diff}& G& VBD.&   $2.80\pm0.70$   &${85.84\pm10.24}$ &$4.10\pm0.43$ &$3.34\pm0.38$ &$3.44\pm0.68$   &$16.00\pm4.69$\\
        $\text{VEIDM}_\text{w}$ \cite{richter_diff}& G& VBD.&   $2.93\pm0.63$   &${86.36\pm9.82}$ &$4.12 \pm 0.68$ &$3.37\pm0.36$   &$3.51\pm 0.67$ &$17.65\pm3.46$ \\
        $\text{VPIDM}_\text{n}$ & G& VBD.&  $\mathbf{3.16}\pm \mathbf{0.69}$   &$\mathbf{87.44}\pm\mathbf{9.44}$ &$\mathbf{4.23} \pm \mathbf{0.66}$  &$\mathbf{3.53}\pm\mathbf{0.53}$  &$\mathbf{3.70} \pm \mathbf{0.71}$ &$18.37\pm3.65$ \\
        $\text{VPIDM}_\text{w}$ & G& VBD.&  ${2.83}\pm {0.52}$   &${87.26}\pm{9.41}$ &${3.95} \pm {0.58}$  &${3.41}\pm{0.45}$  &${3.39} \pm {0.56}$ &$18.00\pm3.30$ \\
        \midrule  
        \midrule	
         Noisy& -& DNS Simu.&  $1.58\pm0.46$   &${80.99\pm12.19}$ &$3.08 \pm 0.74$ &$2.53\pm0.59$ &$2.29\pm0.60$ &$9.07\pm52.48$ \\ 
         MetricGAN+ \cite{fu21_interspeech}& D& DNS Simu.&   $2.18\pm0.56$   &${81.70\pm10.83}$ &$3.12\pm0.70$ &$2.33\pm0.50$ &$2.62\pm0.64$ &$6.30\pm3.37$\\ 
         NCSN++ \cite{song2019generative}& D& DNS Simu.&  $2.25\pm0.67$   &${90.46\pm7.49}$ &$3.14\pm0.98$ &$3.25\pm0.63$ &$2.67\pm0.85$ &$\mathbf{16.70}\pm\mathbf{5.37}$ \\ 
        $\text{VEIDM}_\text{n}$ \cite{richter_diff}& G& DNS Simu.&  $2.34\pm0.68$   &${89.68\pm7.76}$ &$3.64\pm0.75$  &$3.22\pm0.60$  &$2.98\pm0.72$  &$14.52\pm5.57$\\
        $\text{VEIDM}_\text{w}$  \cite{richter_diff}& G& DNS Simu.&   $2.43\pm0.69$   &${90.39\pm7.70}$  &$3.68\pm0.79$  &$3.31\pm0.62$ &$3.06\pm0.75$ &$15.63\pm5.28$\\
        $\text{VPIDM}_\text{n}$ & G& DNS Simu.& $\mathbf{2.52}\pm\mathbf{0.67}$   &$\mathbf{90.91}\pm\mathbf{7.24}$ &$\mathbf{3.73}\pm\mathbf{0.74}$ &$\mathbf{3.35}\pm\mathbf{0.59}$ &$\mathbf{3.12}\pm\mathbf{0.72}$ &$16.15\pm5.00$\\
         $\text{VPIDM}_\text{w}$ & G& DNS Simu.& ${2.41}\pm{0.60}$   &${90.49}\pm{7.09}$ &${3.70}\pm{0.70}$ &${3.29}\pm{0.54}$ &${3.05}\pm{0.66}$ &$16.10\pm3.76$\\
        \bottomrule
		\end{tabular} %
	\end{table*}

\begin{table}[ht]
	
	\caption{The proposed method versus some SOTA
		methods with respect to different metrics.}
	\label{tab:overall}
	\centering
	\resizebox{\columnwidth}{!}{%
	\begin{tabular}{ lcccc}
		\toprule
		{Model} & 
		
		{PESQ $\uparrow$} &
		{CSIG $\uparrow$} &
		{CBAK $\uparrow$} &
		{COVL $\uparrow$}  \\
		\midrule	
        \midrule
           NOISY   &$1.97$  &$3.35$    &$2.44$ &$2.64$  \\
            MetricGAN \cite{fu2019metricgan}   &$2.86$ &$3.99$    &$3.18$ &$3.42$     \\	
		MetricGAN$+$ \cite{fu21_interspeech}   &$3.15$   & $4.14$ &$3.16$ &$3.64$    \\
        \midrule
		CDiffuSE \cite{lu2022conditional}   & $2.52$  &$3.72$    &$2.91$ &$3.10$    \\	
		
		SRTNet \cite{srtnet} & $2.69$ & $4.12$   &$3.19$ &$3.39$    \\
			
		CDSE \cite{cold_diff}   & $2.77$   & $3.91$ &$3.32$ &$3.33$   \\
            VEIDM (2) \cite{richter_diff}  & $2.80$ & $4.10$    &$3.24$ &$3.44$   \\
		
		VPIDM & $\mathbf{3.16}$ & $\mathbf{4.33}$    &$\mathbf{3.53}$ &$\mathbf{3.70}$   \\
		\bottomrule
		
	\end{tabular}%
 }
\end{table}

\bibliographystyle{IEEEtran}
\bibliography{refs}